\documentclass[a4paper,latin9]{aa}
\usepackage[latin9]{inputenc}
\usepackage{amsmath}
\usepackage{amssymb}
\usepackage{wasysym}
\usepackage{graphicx}

\makeatletter

\pdfpageheight\paperheight
\pdfpagewidth\paperwidth

\providecommand{\tabularnewline}{\\}

\usepackage{times}\usepackage{multirow}\usepackage{rotating}\usepackage{longtable}\usepackage{lscape}

\newcommand{\qtir}{$q_{{\rm TIR}}$}
\newcommand{\ltir}{$L_{{\rm TIR}}$}
\newcommand{\lrad}{$L_{1.4{\rm GHz}}$}

\def\f#1   {Fig.~\ref{#1}}
\def\s#1   {Sec.~\ref{#1}}
\def\tab#1   {Tab.~\ref{#1}}
\def\eq#1   {Eq.~\ref{#1}}
\def\t#1   {Tab.~\ref{#1}}

\def\comm#1   {{\tt (COMMENT: #1) }}

\titlerunning{Infrared-radio correlation}
\authorrunning{Delhaize et al.}

\makeatother

\begin{document}

\title{VLA-COSMOS 3\,GHz Large Project: The infrared-radio correlation
of star-forming galaxies and AGN to $z\apprle6$}

\author{J. Delhaize$^{1}$%
\thanks{jacinta@phy.hr%
}, V. Smol{\v{c}}i{\'{c}}$^{1}$, I. Delvecchio$^{1}$, M. Novak$^{1}$,
M. Sargent$^{2}$, N. Baran$^{1}$, B. Magnelli$^{3}$, G. Zamorani$^{4}$,
E. Schinnerer$^{5}$, E. J. Murphy$^{6}$, M. Aravena$^{7}$, S. Berta$^{1}$%
\thanks{Visiting scientist%
}, M. Bondi$^{8}$, P. Capak$^{9}$, C. Carilli$^{10,11}$, P. Ciliegi$^{4}$,
F. Civano$^{12}$, O. Ilbert$^{13}$, A. Karim$^{3}$, C. Laigle$^{14}$,
O. Le Fèvre$^{13}$, S. Marchesi$^{15}$, H. J. McCracken$^{14}$,
M. Salvato$^{16}$, N. Seymour$^{17}$ and L. Tasca$^{13}$}

\institute{Department of Physics, Faculty of Science, University of Zagreb,
Bijeni\v{c}ka cesta 32, 10000 Zagreb, Croatia\and Astronomy Centre,
Department of Physics \& Astronomy, University of Sussex, Brighton,
BN1 9QH, England \and Argelander Institute for Astronomy, University
of Bonn, Auf dem Hügel 71, 53121 Bonn, Germany \and INAF - Osservatorio
Astronomico di Bologna, Via Piero Gobetti 93/3, I-40129 Bologna, Italy
\and Max Planck Institut für Astronomie, Königstuhl 17, 69117 Heidelberg,
Germany. \and National Radio Astronomy Observatory, 520 Edgemont
Rd, Charlottesville, VA 22903, USA \and Núcleo de Astronom\'{i}a,
Facultad de Ingeniería y Ciencias, Universidad Diego Portales, Av.
Ejército 441, Santiago, Chile \and INAF - Istituto di Radioastronomia,
Via Gobetti 101, I-40129 Bologna, Italy \and Department of Astronomy,
California Institute of Technology, MC 249-17, 1200 East California
Blvd, Pasadena, CA 91125, USA. \and National Radio Astronomy Observatory,
Socorro, NM, USA \and Cavendish Laboratory, Cambridge, UK \and Harvard-Smithsonian
Center for Astrophysics, 60 Garden Street, Cambridge, MA 02138, USA
\and Aix Marseille Université, CNRS, LAM (Laboratoire d'Astrophysique
de Marseille) UMR 7326, 13388, Marseille, France \and Institut d'Astrophysique
de Paris, Sorbonne Universités, UPMC Univ Paris 06 et CNRS, UMR 7095,
98 bis bd Arago, 75014 Paris, France \and Department of Physics and
Astronomy, Clemson University, Kinard Lab of Physics, Clemson, SC
29634-0978, USA \and Max-Planck-Institut für Extraterrestrische Physik
(MPE), Postfach 1312, D-85741 Garching, Germany \and International
Centre for Radio Astronomy Research, Curtin University, Perth, WA
6102, Australia}

\abstract{{\normalsize{}{}We examine the behaviour of the infrared-radio correlation
(IRRC) over the range $0<z\lesssim6$ using new, highly sensitive
3\,GHz observations with the Karl G. Jansky Very Large Array (VLA)
and infrared data from the Herschel Space Observatory in the 2\,deg$^{2}$
COSMOS field. We distinguish between objects where emission is believed
to arise solely from star-formation, and those where an active galactic
nucleus (AGN) is thought to be present. We account for non-detections
in the radio or in the infrared using a doubly-censored survival analysis.
We find that the IRRC of star-forming galaxies, quantified by the
infrared-to-1.4\,GHz radio luminosity ratio (\qtir), decreases with
increasing redshift: }$q_{{\rm TIR}}(z)=(2.88\pm0.03)(1+z)^{-0.19\pm0.01}$.
{\normalsize{}{}This is consistent with several previous results
from the literature. Moderate-to-high radiative luminosity AGN do
not follow the same \qtir$(z)$ trend as star-forming galaxies, having
a lower normalisation and steeper decrease with redshift. We cannot
rule out the possibility that unidentified AGN contributions only
to the radio regime may be steepening the observed $q_{{\rm TIR}}(z)$
trend of the star-forming galaxy population. We demonstrate that the
choice of the average radio spectral index directly affects the normalisation,
as well as the derived trend with redshift of the IRRC. An increasing
fractional contribution to the observed 3\,GHz flux by free-free
emission of star-forming galaxies may also affect the derived evolution.
However, we find that the standard (M82-based) assumption of the typical
radio spectral energy distribution (SED) for star-forming galaxies
is inconsistent with our results. This suggests a more complex shape
of the typical radio SED for star-forming galaxies, and that imperfect
$K$ corrections in the radio may govern the derived trend of decreasing
\qtir\ with increasing redshift. A more detailed understanding of
the radio spectrum is therefore required for robust $K$ corrections
in the radio and to fully understand the cosmic evolution of the IRRC.
Lastly, we present a redshift-dependent relation between rest-frame
1.4\,GHz radio luminosity and star formation rate taking the derived
redshift trend into account.}}


\maketitle
\makeatother




\section{Introduction\label{sec:intro}}

A tight correlation between the total infrared luminosity of a galaxy
and its total 1.4\,GHz radio luminosity, extending over at least
three orders of magnitude, has been known to exist for some time (e.g.
\citealt{vanderkruit71,vanderkruit73,dejong85,helou85,condon92,yun01}).
This correlation exists for star-forming late-type galaxies, early-type
galaxies with low levels of star formation and even for some merging
systems (e.g. \citealt{dickey84,helou85,wrobel88,domingue05}).

The so-called infrared-radio correlation (IRRC) has been used to identify
and study radio-loud active galactic nuclei (AGN; e.g.~\citealt{donley05,norris06,park08,delmoro13})
and to estimate the distances and temperatures of high-redshift submillimetre
galaxies (e.g.~\citealt{carilli99,chapman05}). Another important
application of the IRRC is to calibrate radio luminosities for use
as indirect, dust-unbiased star formation rate (SFR) tracers (e.g.~\citealt{condon92,bell03,murphy11,murphy12}).
This is particularly relevant considering the powerful new capabilities
of the recently upgraded radio astronomy facilities (such as the Karl
G. Jansky Very Large Array; VLA) and the next generation of radio
telescopes coming online in the near future (such as MeerKAT, the
Australian SKA Pathfinder and the Square Kilometre Array). Sensitive
radio continuum surveys with these instruments will have simultaneously
good sky coverage and excellent angular resolution and will thus have
the potential to act as powerful SFR tracers at high redshifts. However,
this relies on a proper understanding of whether, and how, the IRRC
evolves with redshift.

Star-formation in galaxies is thought to be responsible for the existence
of the IRRC, although the exact mechanisms and processes at play remain
unclear. Young, massive stars emit ultraviolet (UV) photons, which
are absorbed by dust grains and re-emitted in the infrared (IR), assuming
the interstellar medium is optically-thick at UV wavelengths. After
a few Myr, these massive stars die in supernovae explosions which
produce the relativistic electrons that, diffusing in the galaxy,
are responsible for synchrotron radiation traceable at radio wavelengths
(e.g.~\citealt{condon92}). Several theoretical models attempt to
explain the IRRC on global scales, such as the Calorimetry model proposed
by \citet{voelk89}, the conspiracy model (e.g.~\citealt{bell03,lacki10})
and the optically-thin scenario \citep{helou93}. Models such as the
small-scale dynamo effect (\citealt{schleicher13,niklas97}) attempt
to explain the correlation on more local scales. However, none of
these models successfully reproduce all observational constraints.

As to whether the IRRC evolves with redshift, several different theoretical
predictions exist. \citet{murphy09} predict a gradual increase in
the infrared-to-radio luminosity ratio with increasing redshift due
to inverse Compton scattering off the cosmic microwave background
resulting in reduced synchrotron cooling, although this is dependent
on the magnetic field properties of galaxy populations. \citet{schober16}
model the evolving synchrotron emission of galaxies and also find
a decreasing IRRC towards higher redshifts. On the other hand, \citet{lacki10b}
predict a slight decrease in the infrared-to-radio luminosity ratio
with redshift (of the order of 0.3 dex) by $z\sim2$ due to changing
cosmic ray scale heights of galaxies.

Observationally, a lack of sensitive infrared and/or radio data has,
until recently, restricted the redshift range of studies of the cosmic
evolution of the IRRC. Several observation-based studies have concluded
that the IRRC does not appear to vary over at least the past 10-12\,Gyr
of cosmic history, in that it is linear over luminosity (e.g.~\citealt{sajina08,murphy09b}).
\citet{sargent10a} found no significant evolution in the IRRC out
to $z\sim1.5$ using VLA imaging of the Cosmic Evolution Survey (COSMOS;
\citealt{scoville07}) field at 1.4\,GHz with rms $\sim$15\,$\mu$Jy
\citep{schinnerer07,schinnerer10}. Using a careful survival analysis,
\citet{sargent10a} demonstrate that selecting sources only in the
radio or in the infrared for flux-limited surveys can introduce a
selection bias that can artificially indicate evolution. Several other
studies (e.g.~\citealt{garrett02,appleton04,garn09,jarvis10,mao11,smith14})
have similarly found no significant evidence for evolution of the
IRRC out to $z\sim2$, and out to $z\sim3.5$ by \citet{ibar08}.

More recently, studies of the IRRC evolution towards higher redshifts
have been facilitated by the revolutionary data products provided
by the Herschel Space Observatory \citep{pilbratt10} at far-infrared
wavelengths. For example, \citet{magnelli15} performed a stacking
analysis of Herschel, VLA and Giant Metre-wave Radio Telescope radio
continuum data to study the variation of the IRRC over $0<z<2.3$.
They find a slight, but statistically-significant ($\sim3\sigma$)
evolution of the IRRC. Similarly, \citet{ivison10} find some evidence
for moderate evolution of the IRRC to $z\sim2$ using Herschel and
VLA data, however their sample selection in the mid-infrared may introduce
some bias.

In this paper, we conduct a careful analysis of thousands of galaxies
to examine the IRRC out to $z\sim6$ using deep Herschel observations
of the COSMOS field in combination with the VLA-COSMOS 3\,GHz Large
Project (\citealp{smolcic17a}) - a new, highly sensitive, high-angular
resolution radio continuum survey with the VLA. These are the most
sensitive data currently available over a cosmologically-significant
volume and are thus ideal for such studies. With the wealth of deep,
multiwavelength data (from X-ray to radio) available in the COSMOS
field, we can conduct a sophisticated separation of galaxy populations
into AGN and non-active star-forming galaxies. This allows us to examine
the behaviour of the IRRC for each population separately.

In Section 2 of this paper we describe our data, the construction
of the jointly-selected source sample and the identification of AGN.
In Section 3 we present our analysis of the IRRC as a function of
redshift. In Section 4 we discuss our results with respect to the
literature and examine the various biases involved. We present our
conclusions in Section 5. We assume $H_{0}=70$\,km\,s$^{-1}$\,Mpc$^{-1}$,
$\Omega_{M}=0.3$ and $\Omega_{\Lambda}=0.7$ and a \citet{chabrier03}
initial mass function (IMF), unless otherwise stated. Magnitudes and
colours are expressed in the AB system.

\section{Data}

\subsection{Radio- and infrared- selected samples\label{sub:Radio--and-infrared-samples}}

It has been shown in \citet{sargent10a} that studies using solely
radio-selected or solely IR-selected samples are biased towards low
and high average measurements of the IRRC, respectively, with the
difference (in the ratio of infrared to radio luminosities) being
roughly $0.3$~dex. Therefore, an unbiased study of the IRRC requires
a sample jointly selected in the radio and infrared. This section
details the construction of the radio-selected and infrared-selected
samples and the union of the two, constituting the jointly-selected
sample.

\subsubsection{Radio-selected sample\label{sub:Radio-selected-sample}}

The 3\,GHz COSMOS Large Project survey was conducted over 384 hours
with the VLA between November 2012 and May 2014 in A and C configurations.
The observations, data reduction and source catalogue are fully described
in \citet{smolcic17a}. The data cover the entire 2\,deg$^{2}$ COSMOS
field to an average sensitivity of 2.3\,$\mu$Jy\,beam$^{-1}$ and
an average beamwidth of 0.75$\arcsec$. In total, 10,830 individual
radio components with S/N$\geq5$ have been identified in the field.

We have searched for optical and/or near-IR (hereafter optical) counterparts
to the 8,696 radio sources in regions of the COSMOS field containing
good-quality photometric data (i.e.~the unmasked regions presented
in \citealt{laigle16}). The matching process is identical to that
described in detail in \citet{smolcic17b} and is briefly summarised
here. The best-matching optical counterpart was identified via a position
cross-match with the multi-band COSMOS2015 photometry catalogue of
\citet{laigle16}%
\footnote{An exhaustive list of all available COSMOS multiwavelength data and
enhanced data products (such as photometric and redshift catalogues)
can be found at http://cosmos.astro.caltech.edu/page/astronomers%
} using a search radius of 1.2$\arcsec$. After rejection of objects
with false-match probabilities greater than 20\%, the predicted fraction
of spurious matches is $<1$\% on average \citep{smolcic17b}. We
find optical associations for 7,729 (89\%) of radio sources. These
constitute our photometry-matched radio-selected sample.

\subsubsection{Infrared-selected sample\label{sub:Infrared-selected-sample}}

We use a prior-based catalogue of Herschel-detected objects in the
COSMOS field to construct our infrared-selected sample. The Herschel
Photodetector Array Camera and Spectrometer (PACS) data at 100 and
160$\mu$m are provided by the PACS Evolutionary Probe (PEP; \citealt{lutz11})
survey. The Herschel Spectral and Photometric Imaging Receiver (SPIRE)
data at 250, 350 and 500$\mu$m are available from the Herschel Multi-tier
Extragalactic Survey (HerMES; \citealt{oliver12}). The entire 2\,deg$^{2}$
COSMOS field is fully covered by both surveys.

The use of a prior-based, rather than a blind, Herschel source catalogue
minimises blending issues. The priors come from the 24\,$\mu$m Spitzer
MIPS (\citealt{sanders07,lefloch09}) catalogue of $>60$\,$\mu$Jy
detections, matched to the COSMOS2015 photometric catalogue within
a search radius of $1\arcsec$. A source enters our infrared-selected
sample if a $\mbox{\ensuremath{\geq}}5\sigma$ detection is present
in at least one Herschel band at the position of a prior. We have
chosen to use a $5\sigma$ Herschel detection threshold in order to
match the sensitivity level of the radio data. This will be discussed
further in Section \ref{sub:Survey-sensitivity-comparsion}. See \citet{laigle16}
for a detailed description of the MIPS/COSMOS2015 matching process
and the extraction of fluxes from the PEP and HerMES maps. We find
8,458 such infrared-detected objects with optical COSMOS2015 counterparts
and these constitute our photometry-matched infrared-selected sample.

\subsubsection{Jointly-selected sample\label{sub:joint+diffuse}}

\begin{figure*}
\begin{centering}
\includegraphics[scale=0.37]{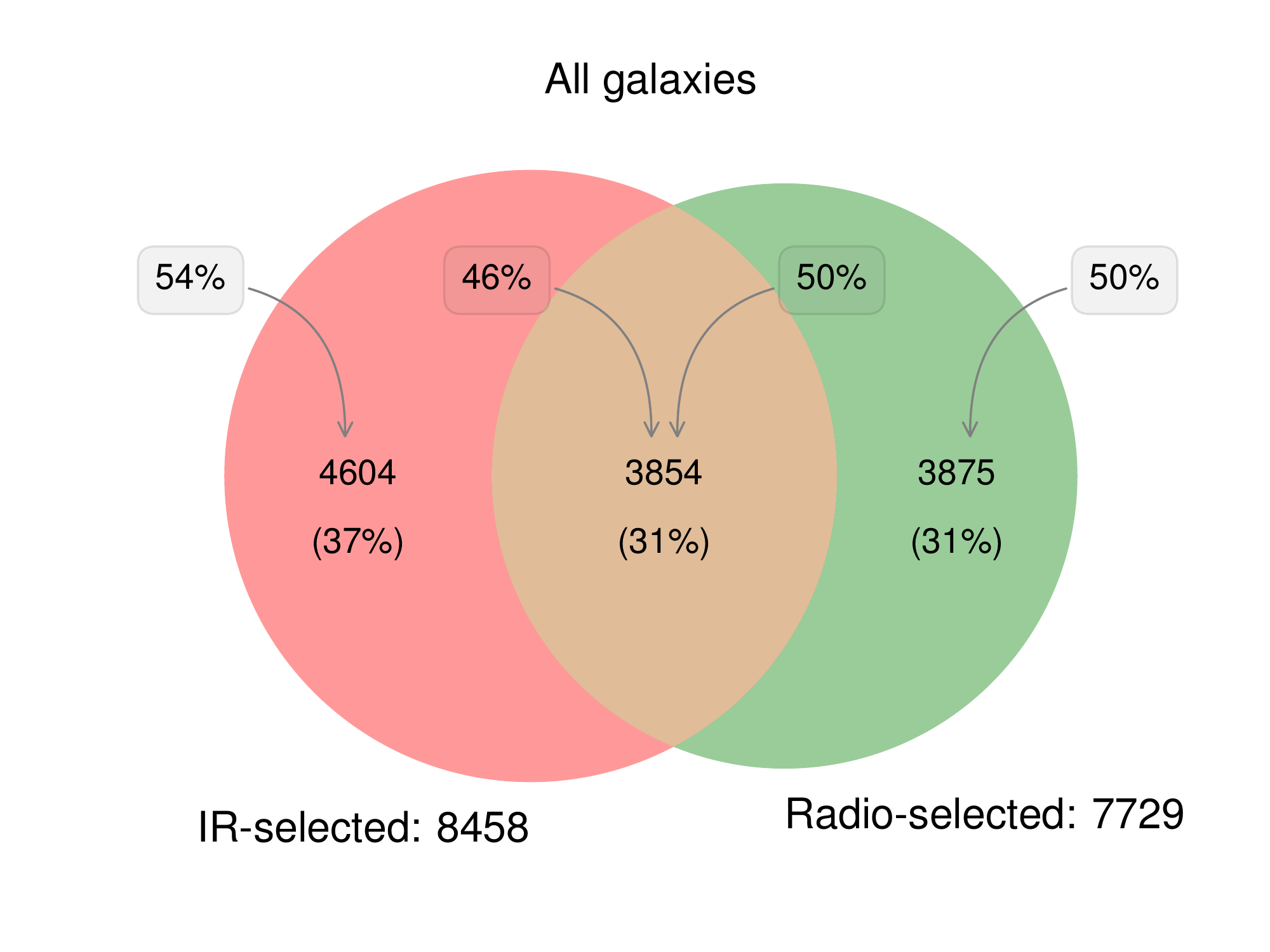}\includegraphics[scale=0.37]{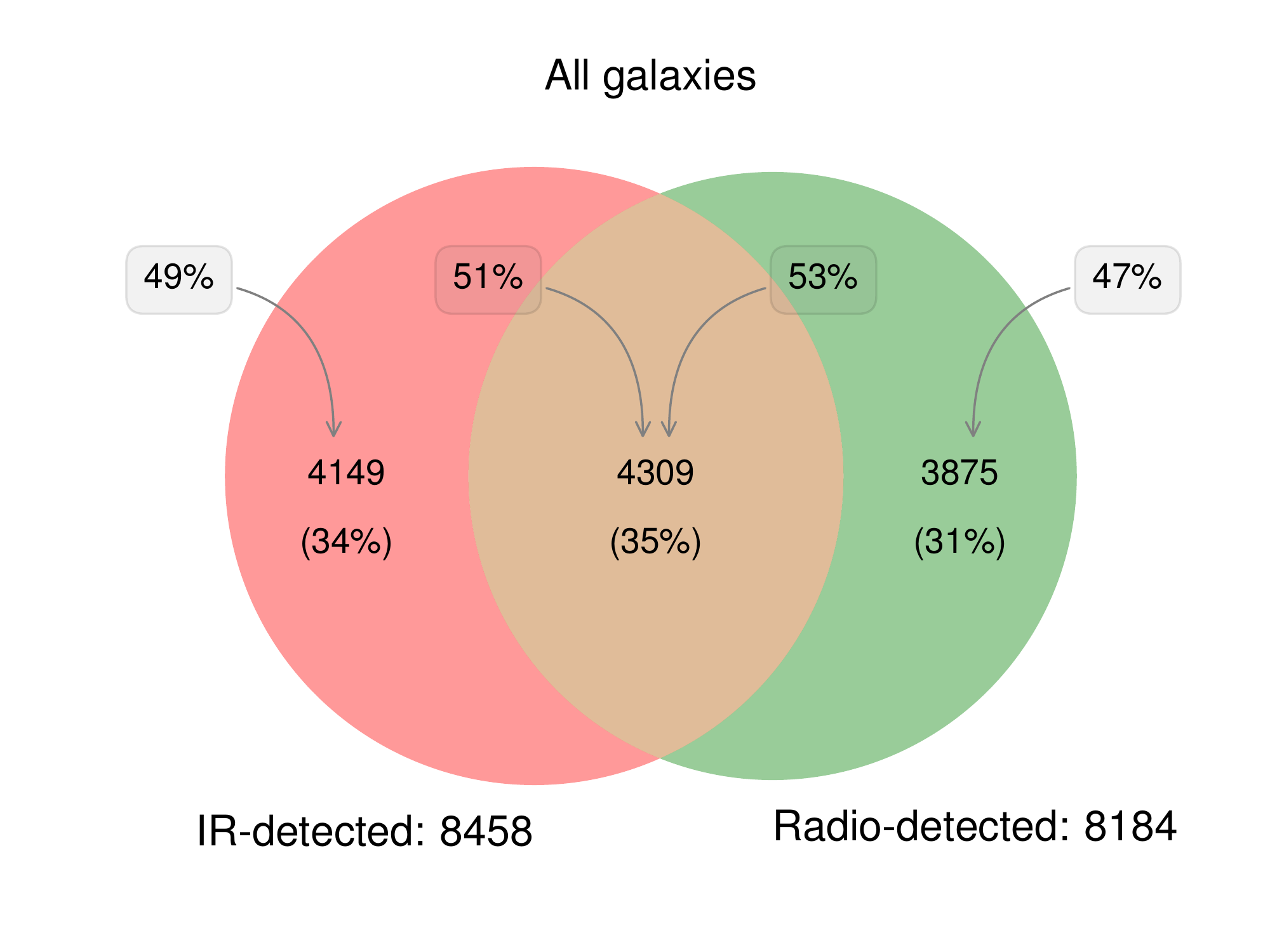} 
\par\end{centering}

\begin{centering}
\includegraphics[scale=0.37]{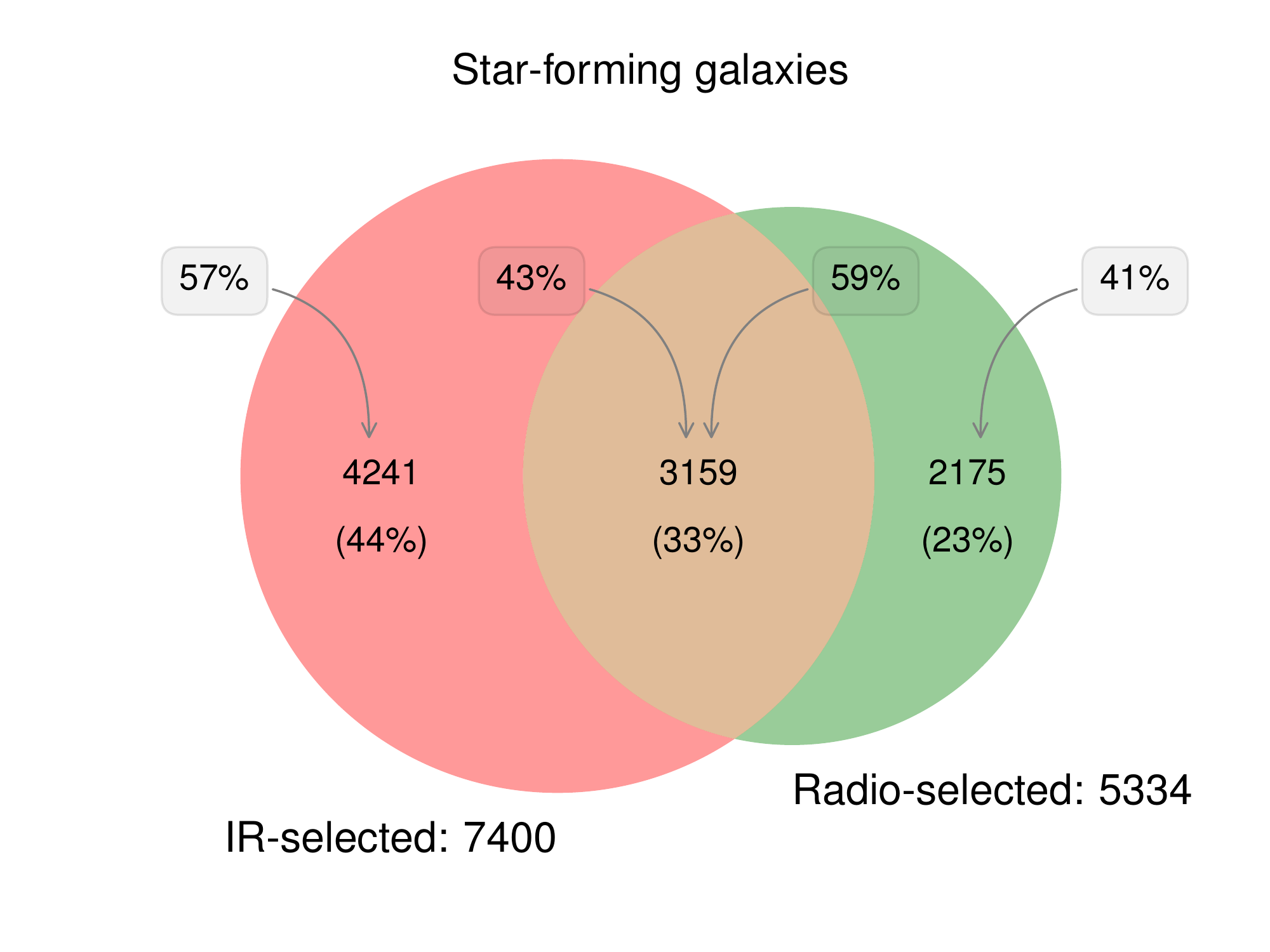}\includegraphics[scale=0.37]{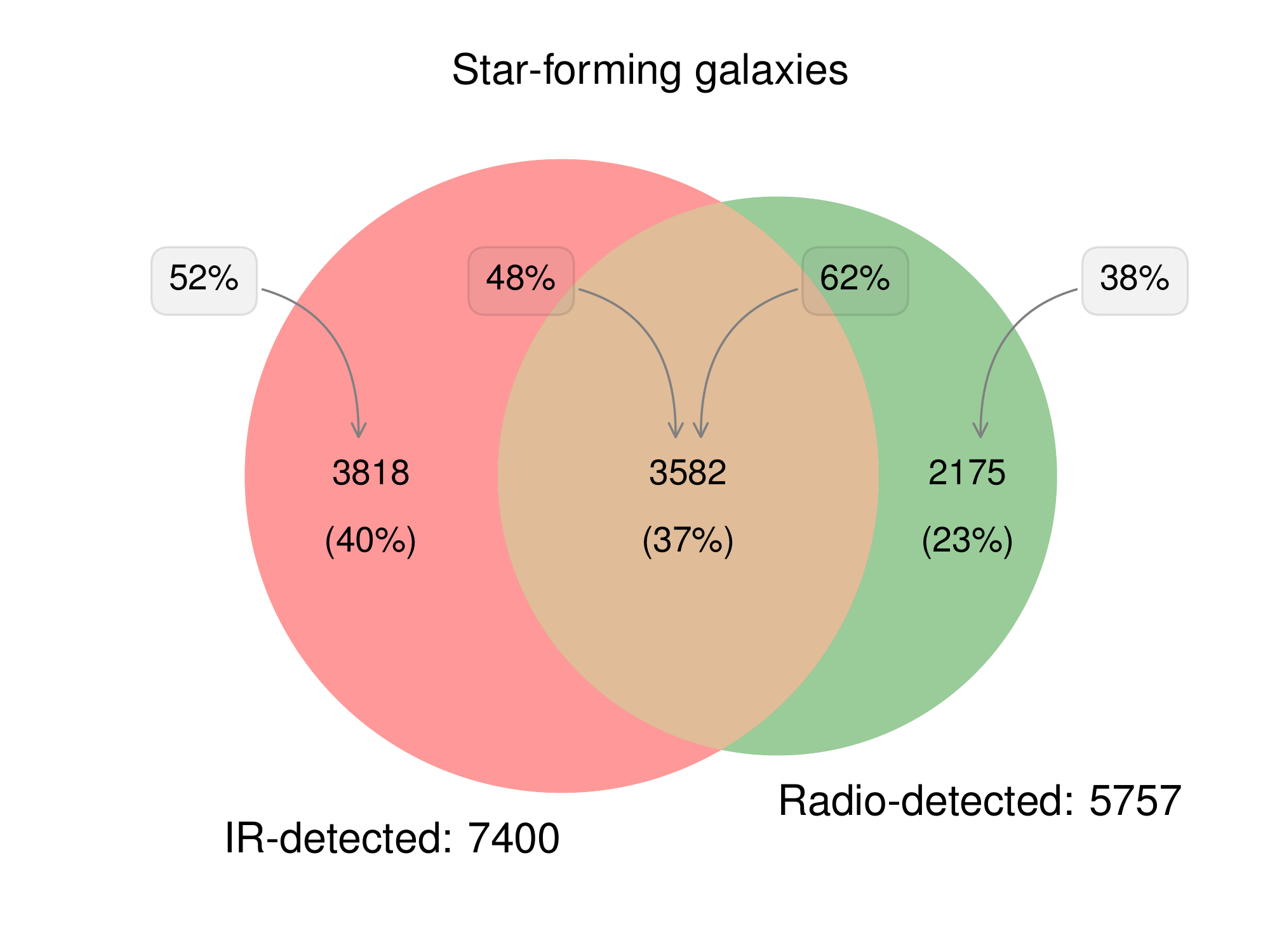} 
\par\end{centering}

\protect\protect\protect\protect\caption{Left: Number and fraction of sources present in the radio-selected
and/or infrared-selected samples for: all objects in the jointly-selected
sample (top), only objects classified as star-forming galaxies (bottom).
The grey boxes to the left (right) in each image show the fractions
relevant to the infrared- (radio-) selected sample only. Right: Same
as for the left but including radio detections identified in convolved
3\,GHz maps. These samples therefore show which objects are detected
(as opposed to selected) in the infrared and/or radio (see Section
\ref{sub:joint+diffuse}).\label{fig:venn}}
\end{figure*}

The jointly-selected sample consists of the union of the radio- and
infrared-selected samples and contains 12,333 sources. As can be seen
in the upper left panel of Figure \ref{fig:venn}, 31\% of objects
are detected in both radio and infrared, 31\% just in radio and 37\%
just in infrared.

Of the radio-selected sample, 50\% are detected in the infrared. However,
it is curious that the majority (54\%) of objects in the infrared-selected
sample, and hence star-forming, are not detected at $\geq5\sigma$
in the radio. This cannot be explained by a difference in the sensitivities
of the Herschel and VLA 3\,GHz data, since the two are comparable,
as will be shown in Section \ref{sub:Survey-sensitivity-comparsion}.

However, this can be partially explained by so-called resolution bias
(see \citealp{smolcic17a}). Extended or diffuse objects may fall
below the detection threshold of the 3\,GHz mosaic due to the high
resolution of the data ($0.75\arcsec$). We have therefore convolved
the 3\,GHz map to several resolutions between $0.75\arcsec$ and
$3.0\arcsec$ and searched for detections in each. This will be discussed
in further detail in Section \ref{sub:3GHz-lims}. Of the 4,604 objects
present in the infrared-selected sample but not present in the radio-selected
sample (i.e.~undetected in the original, unsmoothed 3\,GHz map),
455 are detected at $\geq5\sigma$ in a mosaic of lower resolution.
Hence, 51\% of the infrared-selected sample are detected in the radio.
The final distribution of objects detected in the infrared, radio
or both can be seen in the right-hand panels of Figure \ref{fig:venn}.

\subsubsection{Spectroscopic and photometric redshifts}

We require redshifts for all sources in our jointly-selected sample
in order to conduct spectral energy distribution (SED) fitting and
to compute luminosities. For 35\% (4,354) of optical counterparts,
highly-reliable spectroscopic redshifts are available in the COSMOS
spectroscopic redshift master catalogue (Salvato et al.~in prep),
with redshifts coming mainly from the zCOSMOS survey \citep{lilly07},
DEIMOS runs (Capak et al.~in prep), and the VUDS survey \citep{lefevre15,tasca16}.
Photometric redshifts were available for the remaining sources. For
7,607 objects, these are taken from the COSMOS2015 photometric redshift
catalogue of \citet{laigle16} and were generated using \textsc{lephare}
SED fitting \citep{ilbert13}. The remaining 372 objects have X-ray
counterparts and for these it is more appropriate to use the photometric
redshifts produced via \textsc{lephare} SED fitting incorporating
AGN templates \citep{salvato09,salvato11}.

\subsection{Identification and exclusion of AGN\label{sub:Galaxy-classification}}

\begin{table*}
\protect\caption{Number of objects in the jointly-selected sample within each galaxy
type classification. \label{tab:galaxy_type}}

\begin{centering}
\begin{tabular}{cccccc|c|c}
 &  & Optical colour  & IR-detected  & Radio-detected  & IR \& Radio  & Total  & Total\tabularnewline
 &  & selection  & only  & only  & detected  &  & \tabularnewline
\hline 
\multirow{3}{*}{Star-forming}  & Blue  & $M_{{\rm NUV}}-M_{r}<1.2$  & 2490  & 1392  & 2331  & 6213  & \tabularnewline
 & Green  & $1.2<M_{{\rm NUV}}-M_{r}<3.5$  & 1228  & 783  & 1150  & 3161  & 9575\tabularnewline
 & Red  & $M_{{\rm NUV}}-M_{r}>3.5$  & 100  & -  & 101  & 201  & \tabularnewline
\hline 
\multirow{2}{*}{AGN}  & \textit{HLAGN}  & \textit{N/A}  & \textit{331}  & \textit{909}  & \textit{727}  & \textit{1967}  & \textit{\multirow{2}{*}{2758}}\tabularnewline
 & \textit{MLAGN}  & $M_{{\rm NUV}}-M_{r}>3.5$  & \textit{-}  & \textit{791}  & \textit{- }  & \textit{791}  & \tabularnewline
\hline 
Total  &  &  & 4149  & 3875  & 4309  & 12333  & 12333\tabularnewline
\end{tabular}
\par\end{centering}

\tablefoot{The number of objects which are present only in the infrared-detected
sample are shown in Column 4, those present only in the radio-detected
sample in Column 5 and those present in both in Column 6. The total
in each class is also shown in Column 8. Subsets in italics are considered
AGN and are excluded from the star-forming sample.} 
\end{table*}

We wish to consider the relationship between infrared and radio properties
due solely to star-formation. Therefore, we identify galaxies likely
to host AGN and exclude them from our sample. We exclude a source
if it displays evidence of radiatively-efficient AGN emission based
on the following criteria:

(i) it displays power-law like emission in the mid-infrared and the
IRAC colours satisfy the criteria of \citet{donley12} to predict
the presence of a dusty AGN torus (as in \citealp{smolcic17b}), and/or

(ii) it has an X-ray counterpart detected in the combined Chandra-COSMOS
and COSMOS Legacy surveys (\citealt{elvis09,civano12,civano16,marchesi16})
with a full intrinsic ({[}0.5-8{]}\,keV) X-ray luminosity $L_{X}>10^{42}$\,erg\,s$^{-1}$
(as in \citealp{smolcic17b}), and/or

(iii) When fitting the SED of the object using both a purely star-forming
template and a separate AGN template, the AGN component of the SED
is found to be significant based on a Fisher test (\citealt{delvecchio14}).
This multi-component SED fitting process is conducted using \textsc{sed3fit}%
\footnote{The multi-component SED fitting code \textsc{sed3fit} is described
in \citet{berta13} and is publicly available from http://cosmos.astro.caltech.edu/page/other-tools%
} (\citealt{berta13}) and is discussed in detail in \citet{delvecchio17}.

Using these three criteria, we identify 1,967 objects from the jointly-selected
sample as likely AGN. We refer to these objects as moderate-to-high
radiative luminosity AGN (HLAGN). A discussion of this nomenclature
can be found in \citet{smolcic17b} and \citet{delvecchio17}, the
latter of which also provides a discussion of the relative fraction
of AGN identified by each criterion and the extent of overlap.

We further identify an object as an AGN and exclude it from our sample
if it does not appear in the IR-selected sample (and thus displays
no evidence of appreciable star-formation activity), displays red
optical rest-frame colours $(M_{{\rm NUV}}-M_{r})>3.5$ (and is hence
considered `passive' in the classification scheme of \citealt{ilbert09})
and is radio-detected (i.e.~present in the radio-selected sample).
The colour-selection method is described in detail in \citet{smolcic17b}
and $(M_{{\rm NUV}}-M_{r}$) colours are defined in the COSMOS2015
catalogue \citep{laigle16}. Considering the lack of observed star
formation, the majority of the radio synchrotron emission in such
sources is expected to arise from AGN processes. These objects are
likely to be low-to-moderate radiative luminosity AGN (MLAGN hereafter),
sometimes referred to as low-excitation radio galaxies (LERGs; e.g.
\citealt{sadler02,best05}). We note that these objects are referred
to as quiescent MLAGN in \citet{smolcic17b}. We find 791 such objects.

\begin{figure}
\includegraphics[scale=0.45]{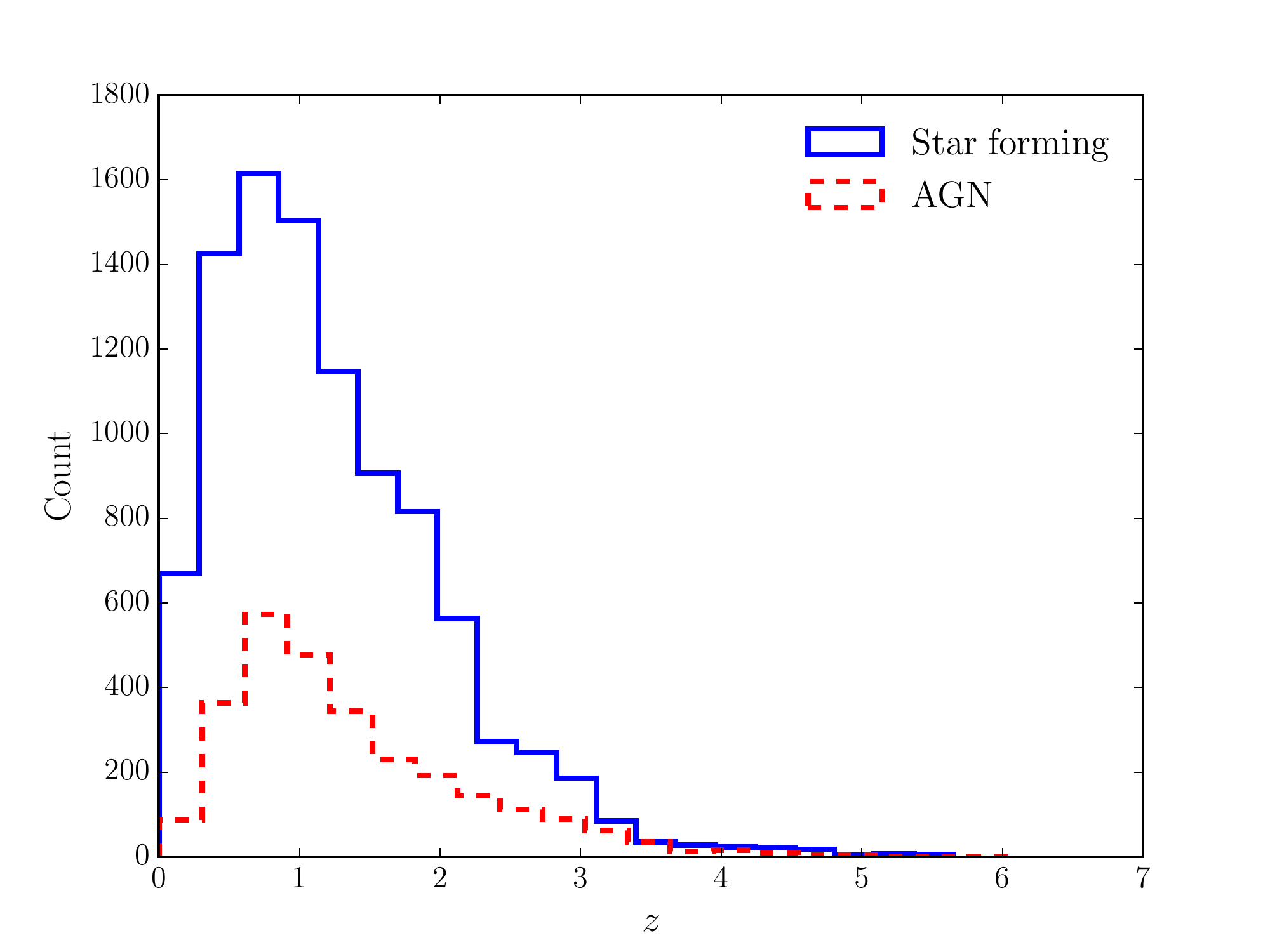}

\protect\protect\protect\protect\caption{Redshift distribution of the star-forming population (blue solid line)
and AGN population (red dashed line) in the jointly-selected sample.\label{fig:The-redshift-distribution}}
\end{figure}

The remaining 9,575 sources in the jointly-selected sample display
no evidence of AGN presence and we therefore consider their infrared
and radio emission to arise predominantly from star-formation. The
distribution of these between the infrared- and radio-selected samples
can be seen in the lower panel of Figure \ref{fig:venn}. A summary
of the classification of all objects in the jointly-selected sample
is presented in Table \ref{tab:galaxy_type}. Figure \ref{fig:The-redshift-distribution}
shows the redshift distribution of the star-forming and AGN populations
separately. The median redshifts of the star-forming and AGN samples
are 1.02 and 1.14, respectively.

All further analysis will focus solely on the star-forming population,
unless otherwise stated.

\subsection{Radio and infrared luminosities\label{sub:luminosities}}

\subsubsection{Radio spectral indices and 1.4\,GHz luminosities\label{sub:Radio-lums}}

\begin{figure}
\includegraphics[scale=0.45]{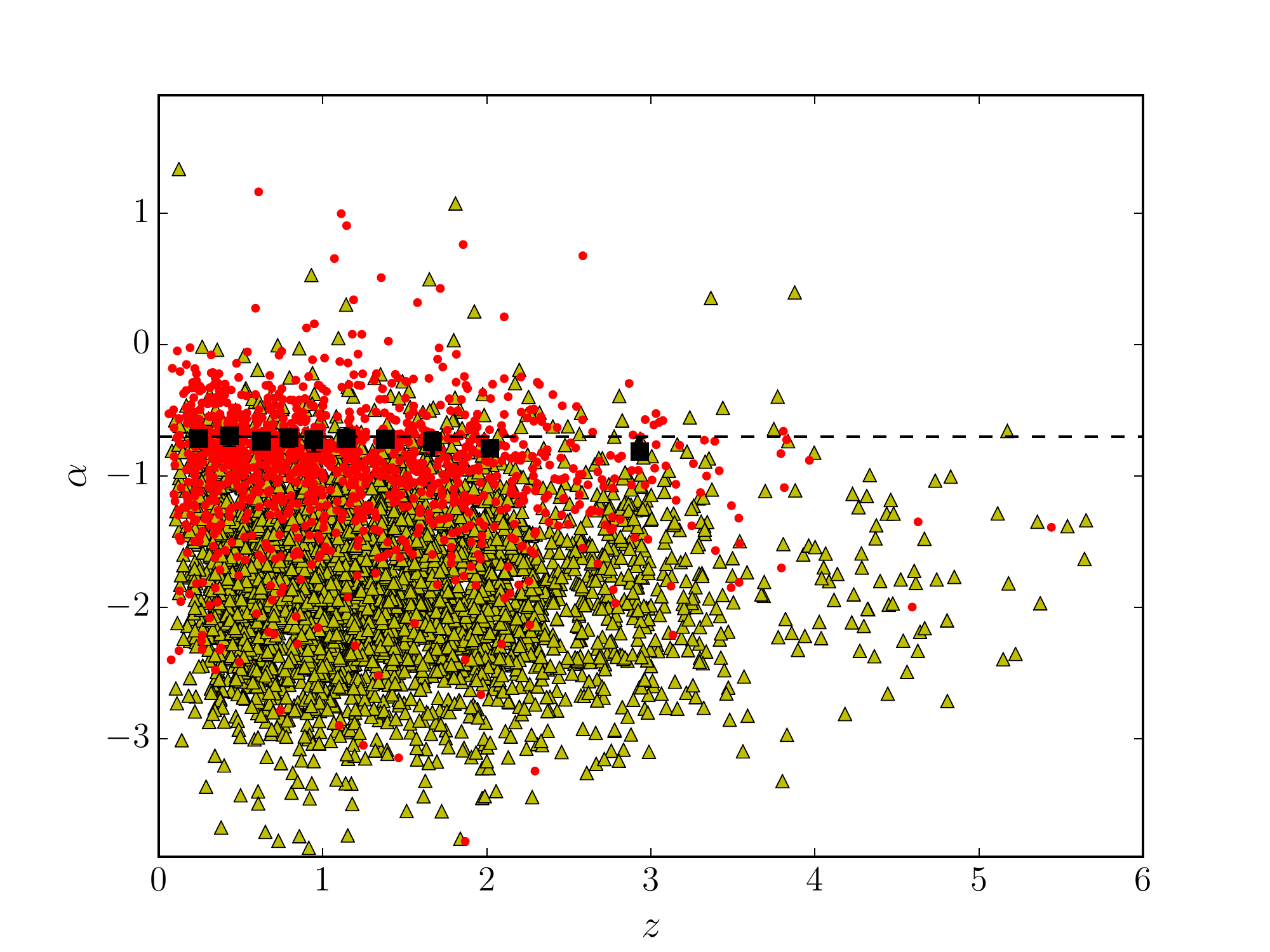}

\protect\protect\protect\protect\caption{3\,GHz to 1.4\,GHz spectral indices ($\alpha_{1.4{\rm GHz}}^{3{\rm GHz}}$)
of the star-forming population as a function of redshift. Red points
show direct measurements, while yellow triangles show 5$\sigma$ lower
limits for objects not detected at 1.4\,GHz. The median within redshift
bins are shown by black squares and have been calculated using a single-censored
survival analysis, which incorporates the lower limits. The median
$\alpha_{1.4{\rm GHz}}^{3{\rm GHz}}$ of the star-forming population
is consistent with $\alpha_{1.4{\rm GHz}}^{3{\rm GHz}}=-0.7$ (indicated
by the horizontal dashed line), at least at $z\lesssim2$.\label{fig:alpha-vs-z}}
\end{figure}

We calculate the spectral index ($\alpha$, where $S_{\nu}\propto\nu^{\alpha}$)
of radio sources by comparing the 3\,GHz fluxes to those in the 1.4\,GHz
VLA COSMOS data \citep{schinnerer04,schinnerer07,schinnerer10}. Of
star-forming objects in the radio-selected sample, 1,212 (23\%) are
detected in both the 3\,GHz map and the shallower 1.4\,GHz map.
Figure \ref{fig:alpha-vs-z} shows the individual measured spectral
indices for these objects. The 5$\sigma$ lower limit on the spectral
index is also shown for all 3\,GHz-detected objects without detections
at 1.4\,GHz. We use a single-censored survival analysis to calculate
the median value of $\alpha_{1.4{\rm GHz}}^{3{\rm GHz}}$ within several
redshift bins. See Section \ref{sub:qtir} for details on the binning
process. This uses the Kaplan-Meier estimator to incorporate the lower
limits when computing the median (Kaplan \& Meier, 1958). As seen
in Figure \ref{fig:alpha-vs-z}, no evolution of the spectral index
with redshift is evident. The median in redshift bins at\textbf{ $z<2.0$}
are consistent with $\alpha_{1.4{\rm GHz}}^{3{\rm GHz}}=-0.7$, and
is also consistent with that found for all objects in the full 3\,GHz
source catalogue \citep{smolcic17a}. In the two $z>2$ bins, the
median spectral index is more consistent with $\alpha=-0.8$ (see
also Figure \ref{fig:alpha} in Section \ref{sub:discussion-Lrad}).
For simplicity, we assume $\alpha=-0.7$ for all objects undetected
at 1.4\,GHz, however we examine the impact of a particular choice
of spectral index on the results in Section \ref{sub:discussion-Lrad}.
We note that the use of $\alpha=-0.7$ predicts a 1.4\,GHz flux that
is inconsistent with the 1.4\,GHz limit in only 3\% of cases.

We convert the observer-frame 3\,GHz fluxes ($S_{3{\rm GHz}}$; W\,Hz$^{-1}$\,m$^{-2}$)
into 1.4\,GHz luminosities (\lrad; W\,Hz$^{-1}$) via:

\begin{center}
\begin{equation}
L_{1.4{\rm GHz}}=\frac{4\pi D_{L}^{2}}{(1+z)^{\alpha+1}}\left(\frac{1.4}{3}\right)^{\alpha}S_{3{\rm GHz}},\label{eq:Lrad}
\end{equation}

\par\end{center}

\noindent where $D_{L}$ is the luminosity distance to the object
in metres.

\noindent For any object with no $\geq5\sigma$ detection in the original
3\,GHz mosaic, \lrad\ is calculated by replacing $S_{3{\rm GHz}}$
in Equation \ref{eq:Lrad} by the flux measured from a lower resolution
3\,GHz mosaic, or by the $5\sigma$ 3\,GHz flux upper limit. The
following section will describe how such fluxes and flux limits are
determined.

\noindent 

\subsubsection{3\,GHz detections and flux limits from convolved mosaics\label{sub:3GHz-lims}}

\begin{figure}
\includegraphics[scale=0.45]{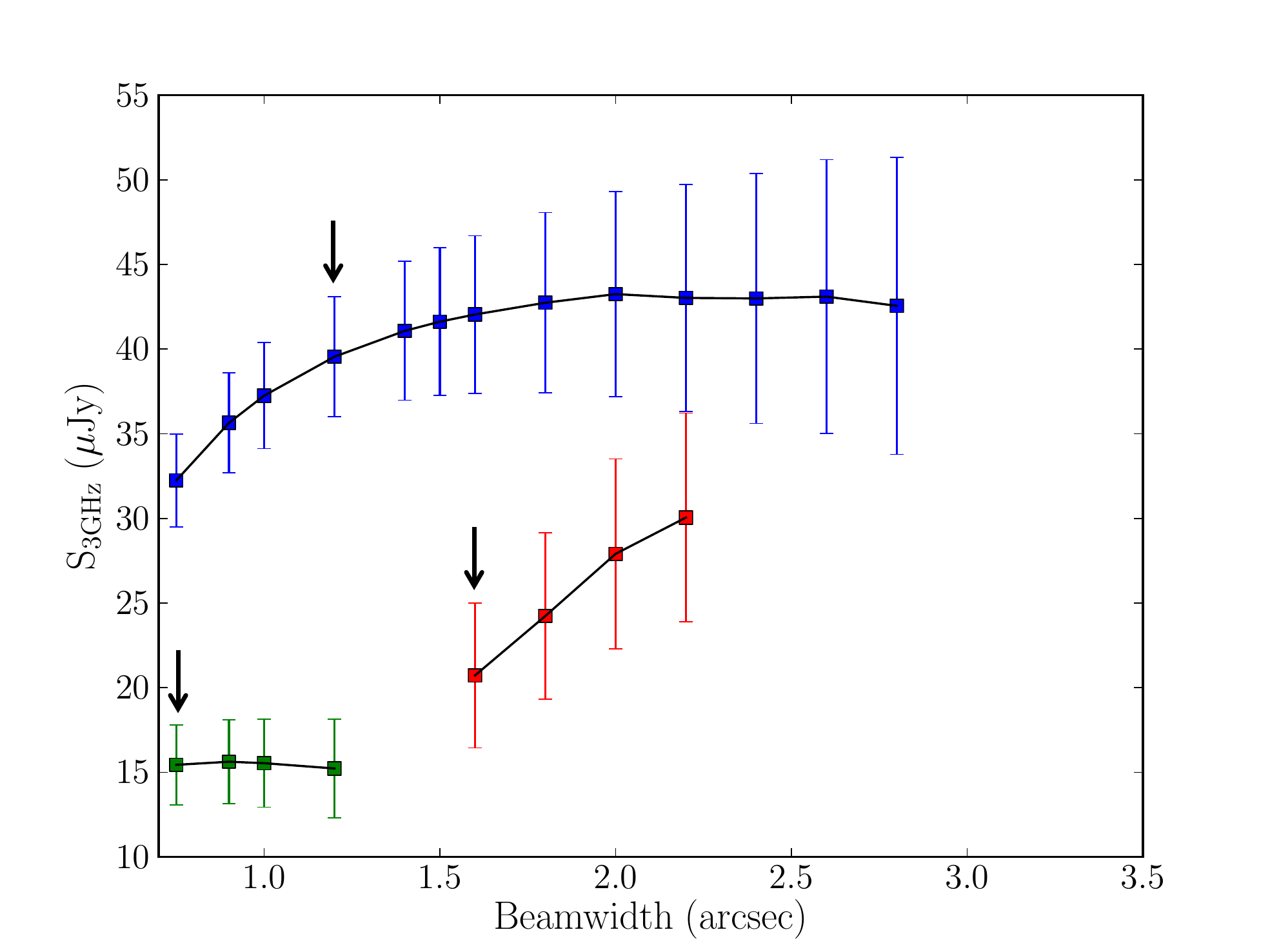}

\protect\protect\protect\caption{3\,GHz peak flux measured in each convolved 3\,GHz map for three
different objects with 1\,$\sigma$ error bars. Fluxes associated
with a given object are colour-coded and joined by a line. In each
case, the indicated point shows where the flux ceases to change significantly
with increasing convolution. The mosaic resolution at which this occurs
is considered optimal. See text (Section \ref{sub:3GHz-lims}) for
further details.\textbf{\label{fig:S3_peak-vs-BW}}}
\end{figure}

As introduced in Section \ref{sub:joint+diffuse}, the high resolution
of the 3\,GHz data ($0.75''$) means that extended and/or diffuse
emission may fall below the detection threshold of the mosaic, corresponding
to a peak flux density of five times the local rms. It is therefore
possible that some 3\,GHz counterparts to objects in the infrared-selected
sample are missed, particularly at low redshift. To overcome this
issue, we convolve the 3\,GHz map to several resolutions between
$0.75''$ and $3.0''$ (i.e.~between one and four times the original
beamwidth). The convolution increases the average rms of the map,
but allows for the potential detection of sources with extended radio
emission but missed in the $\mbox{0.75\arcsec}$ mosaic.

If an infrared-detected object is not detected at $\geq5\sigma$ in
the original $0.75\arcsec$ resolution radio mosaic, there are two
possibilities:

(i) the object is detected at $\geq5\sigma$ in one or more convolved
radio mosaics, or

(ii) the object remains undetected in all convolved radio mosaics.

We calculate the $S_{3{\rm GHz}}$ measurement (or limit) differently
for each of these two cases, as follows.

For case (i), we use the integrated flux density from the 3\,GHz
mosaic with the highest resolution (i.e.~smallest beamwidth) where
the object is detected at $\geq5\sigma$. It is appropriate to use
the integrated flux density since it is found to be stable with changing
resolution, while the peak flux would be underestimated for such extended
sources. Table \ref{tab:convolvedN} shows the number of sources per
mosaic from which the flux measurement is taken. As mentioned in Section
\ref{sub:joint+diffuse}, 3\,GHz counterparts to an additional 455
(5\% of) infrared-detected objects are found in lower resolution mosaics.

While we are justified in using the measured 3\,GHz flux for these
455 objects with prior positions in the infrared (Herschel and 24\,$\mu$m),
we do not allow the additional objects detected in convolved 3\,GHz
maps to enter our radio-selected sample. This would result in a highly
incomplete sample due to the significantly changing rms with increasing
level of convolution and would require additional complex completeness
and false detection rate tests (see \citealp{smolcic17a}) which are
beyond the scope of this paper.

\begin{table}
\protect\caption{Resolution (i.e.~beamwidth) of each convolved 3\,GHz mosaic, the
average rms and the number of sources for which the 5\,$\sigma$
flux measurement is taken from that particular mosaic.\label{tab:convolvedN} }

\begin{tabular}{ccc}
Mosaic resolution  & $\left\langle {\rm rms}\right\rangle $ ($\mu$Jy)  & $N$ (detected) \tabularnewline
\hline 
0.75$\arcsec$ (original)  & 2.34  & 7729\tabularnewline
0.9$\arcsec$  & 2.5  & 199\tabularnewline
1.0$\arcsec$  & 2.66  & 89\tabularnewline
1.2$\arcsec$  & 3.08  & 80\tabularnewline
1.4$\arcsec$  & 3.57  & 29\tabularnewline
1.5$\arcsec$  & 3.84  & 14\tabularnewline
1.6$\arcsec$  & 4.13  & 11\tabularnewline
1.8$\arcsec$  & 4.77  & 11\tabularnewline
2.0$\arcsec$  & 5.49  & 9\tabularnewline
2.2$\arcsec$  & 6.04  & 7\tabularnewline
2.4$\arcsec$  & 6.73  & 3\tabularnewline
2.6$\arcsec$  & 7.64  & 1 \tabularnewline
2.8$\arcsec$  & 8.32  & 1 \tabularnewline
$3.0\arcsec$  & 9.13  & 1 \tabularnewline
\hline 
\end{tabular}
\end{table}

For case (ii), the $5\sigma$ 3\,GHz flux limit is taken as five
times the value at the corresponding pixel position in the noise map
associated with the most appropriate convolved mosaic. The most appropriate
mosaic is chosen as follows. For all sources in a given redshift bin,
which are detected at $\geq5\sigma$ in at least one 3\,GHz mosaic
(i.e.~any object in the radio-selected sample or satisfying case
(i)), we track how the peak flux (surface brightness) changes with
the level of convolution. Several examples are shown in Figure \ref{fig:S3_peak-vs-BW}.
For each source, the optimal map resolution is that where the peak
flux ceases to change significantly with increased convolution. i.e.~the
first data point which is inconsistent (considering the 1\,$\sigma$
errors) with the native point (the measurement from the highest resolution
map) but is consistent with all data points in lower resolution maps.
This is considered to be the resolution at which all emission from
the source is contained within a single map pixel.

\begin{figure}
\includegraphics[scale=0.45]{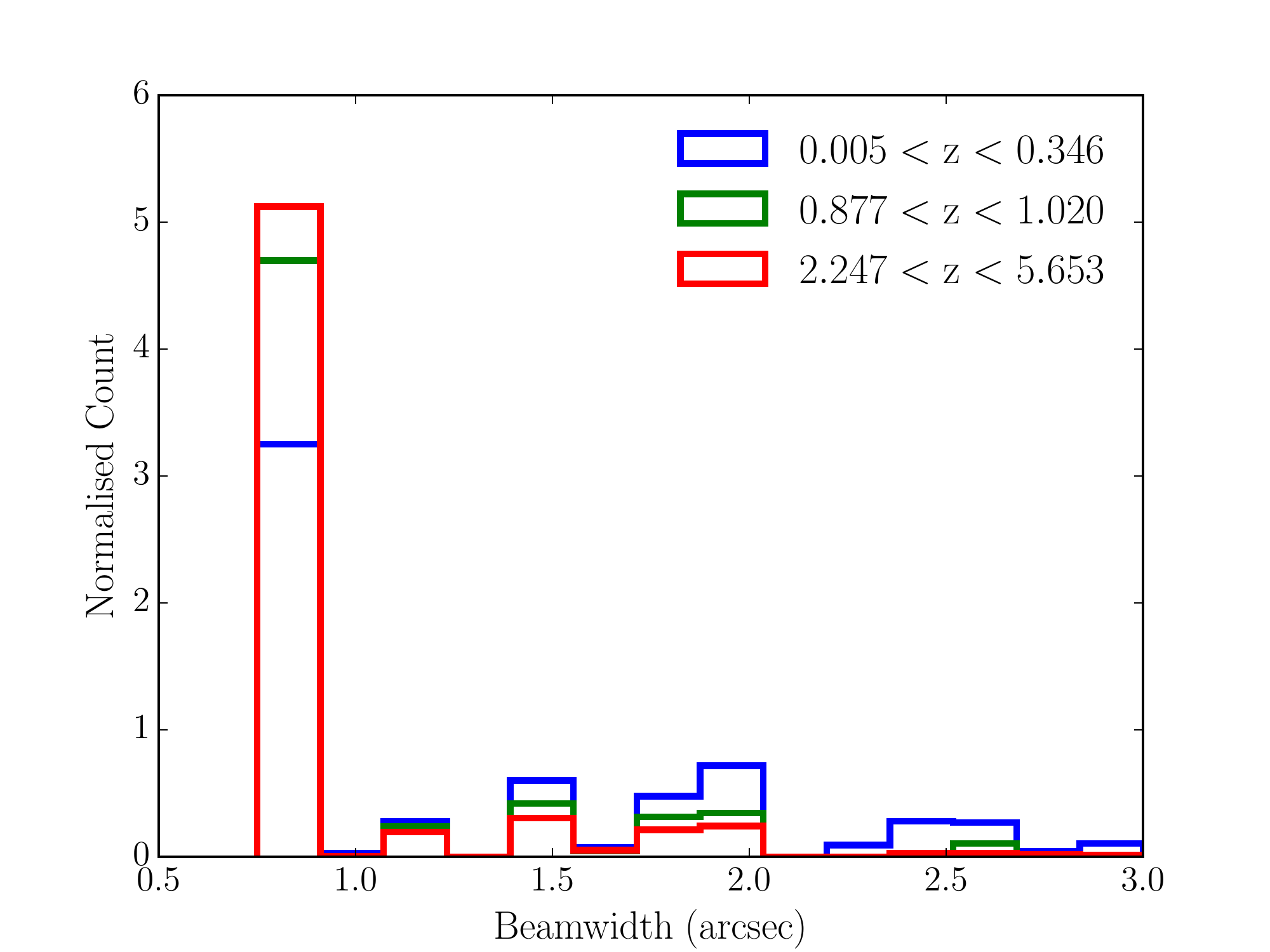}

\protect\protect\protect\protect\caption{Normalised distribution of optimal convolved mosaic resolutions for
3\,GHz detections in a given redshift bin. For clarity, this is shown
only for three redshift bins. See text (Section \ref{sub:3GHz-lims})
for explanation of how the optimal resolution is chosen. \label{fig:convol-cdf}}
\end{figure}

For a given object undetected at 3\,GHz, the mosaic from which to
calculate the 3\,GHz flux limit is chosen by sampling from the distribution
of optimal resolutions in that redshift bin using a Monte-Carlo-like
approach. Examples of the sampled distributions are shown in Figure
\ref{fig:convol-cdf}. The average rms of each convolved map is reported
in Table \ref{tab:convolvedN}. This technique for determining 3\,GHz
upper limits overcomes much of the resolution bias in our data.

\subsubsection{Infrared luminosities \label{sub:Infrared-luminosities}}

The total infrared luminosities ($L_{{\rm TIR}}$) of each source
in the jointly-selected sample are found by integrating the best-fitting
galaxy template to the SED between $8-1000$\,$\mu$m in rest-frame.
The data available over this range in the full COSMOS field include
Spitzer MIPS 24\,$\mu$m data and the five Herschel PACS and SPIRE
bands. For 95 star-forming galaxies, sub-millimetre data was also
available from various instruments including AzTEC and ALMA (\citealt{casey13,scott08,aretxaga11,bertoldi07,smolcic12,miettinen15};
Aravena et al. in prep). The SED fitting to the COSMOS multiband photometry
was conducted using \textsc{magphys} \citep{dacunha08} and is presented
in \citet{delvecchio17}.

As discussed in Section \ref{sub:Infrared-selected-sample}, we require
a $\geq5\sigma$ detection for a source to enter the infrared-selected
sample. This is for the purpose of sensitivity matching with the radio.
Of the star-forming galaxies in the infrared-selected sample, 53\%
of objects are detected at $\geq5\sigma$ in only one Herschel band,
while 1\% are detected in all bands. However, catalogued infrared
photometry is also available for $3\leq{\rm S/N}\leq5$ objects. We
use this photometry for SED fitting where it is available as it provides
better constraints compared with the use of a limit. We have confirmed
that this does not result in any bias towards higher luminosities
due to noise-induced flux boosting at the faint flux end.

If a source has S/N$<3$ in a particular Herschel band, we constrain
the SED fit using the corresponding 3\,$\sigma$ upper limit to the
flux. A single value for this limit is used for each band, and full
details of this process are provided in Section 3 of \citet{delvecchio17}.

In cases where the source is undetected at $\geq5\sigma$ in all Herschel
bands, integrating the resulting best-fit SED provides only an upper
limit on the $L_{{\rm TIR}}$. This is the case for the 2,175 star-forming
objects not in the infrared-selected sample. However, we note that
the SED template fit, and therefore the \ltir\ limit, will still
be somewhat constrained in the infrared regime since a 24\,$\mu$m
detection is available in 59\% of such cases and also due to the optical/infrared
energy balance performed by \textsc{magphys} \citep{dacunha08}.

\begin{figure}
\includegraphics[scale=0.45]{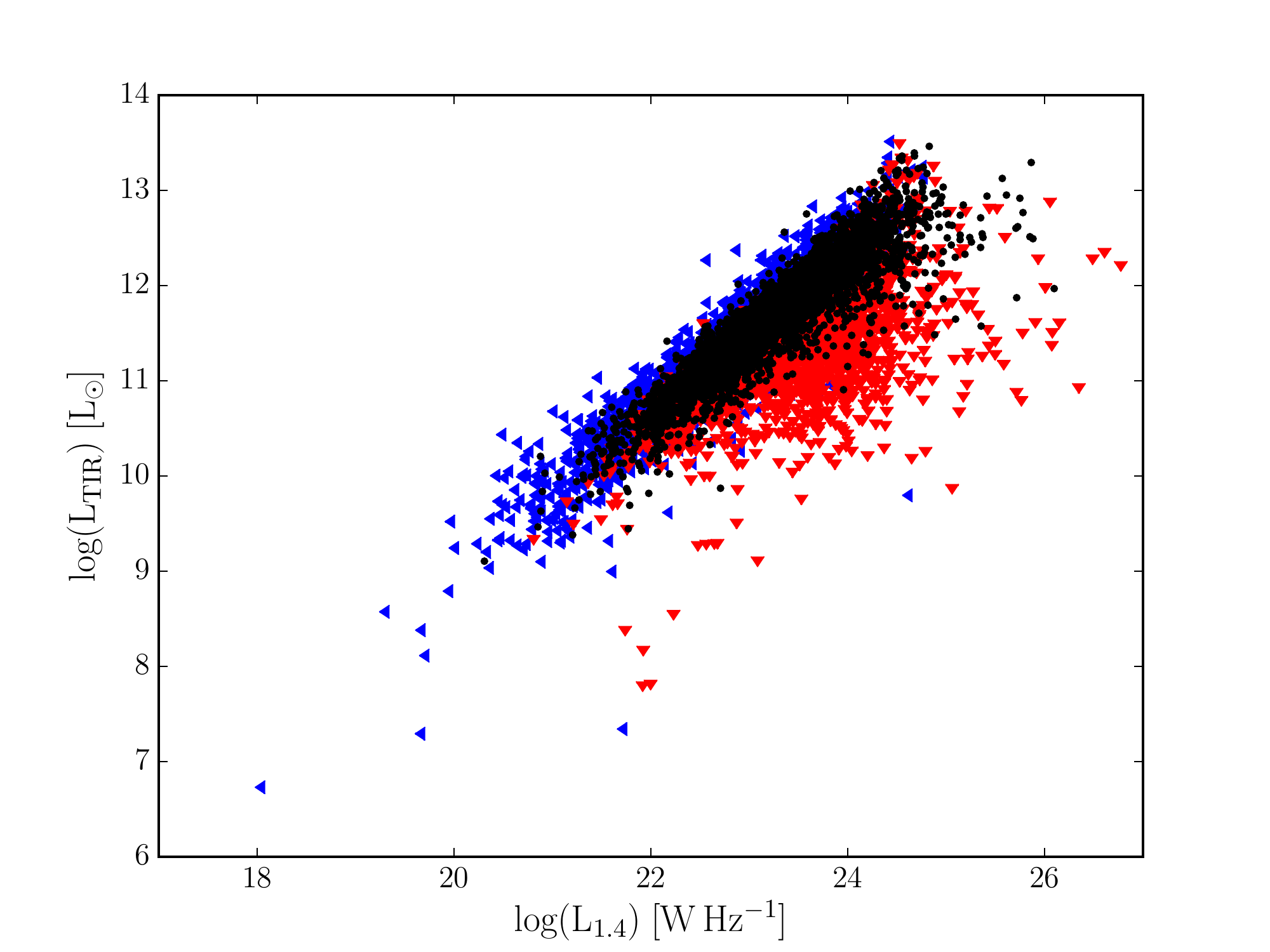}

\protect\protect\protect\protect\caption{Total infrared versus 1.4\,GHz luminosity for star-forming objects
in the jointly-selected sample. Black points show objects directly
detected in both the radio and infrared data, red arrows indicate
objects in the radio-detected sample with upper limits in the infrared
and blue arrows indicate objects in the infrared-detected sample with
upper limits in the radio.\label{fig:Radio-vs-infrared-lum}}
\end{figure}

\begin{figure*}
\includegraphics[scale=0.45]{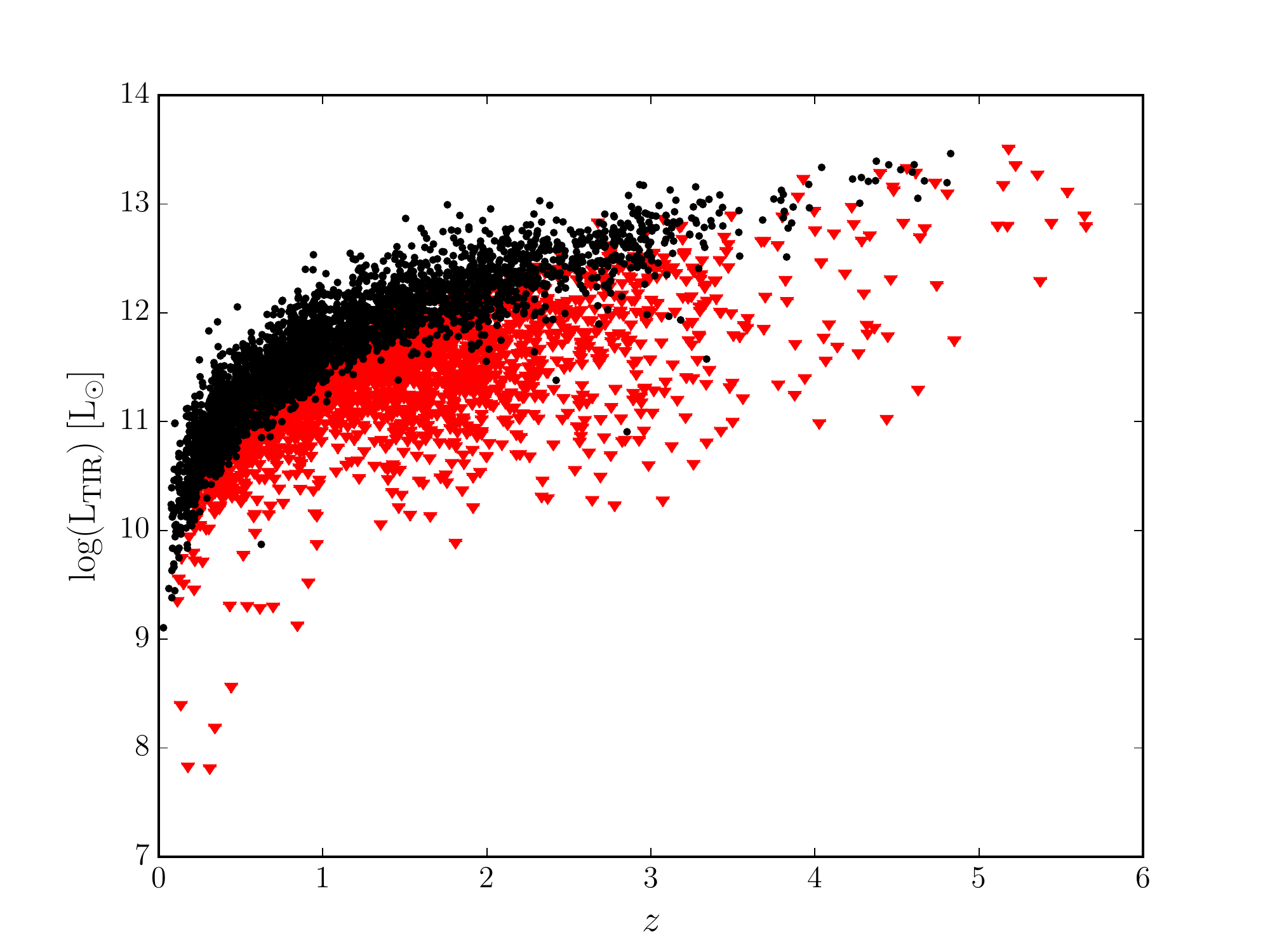}\includegraphics[scale=0.45]{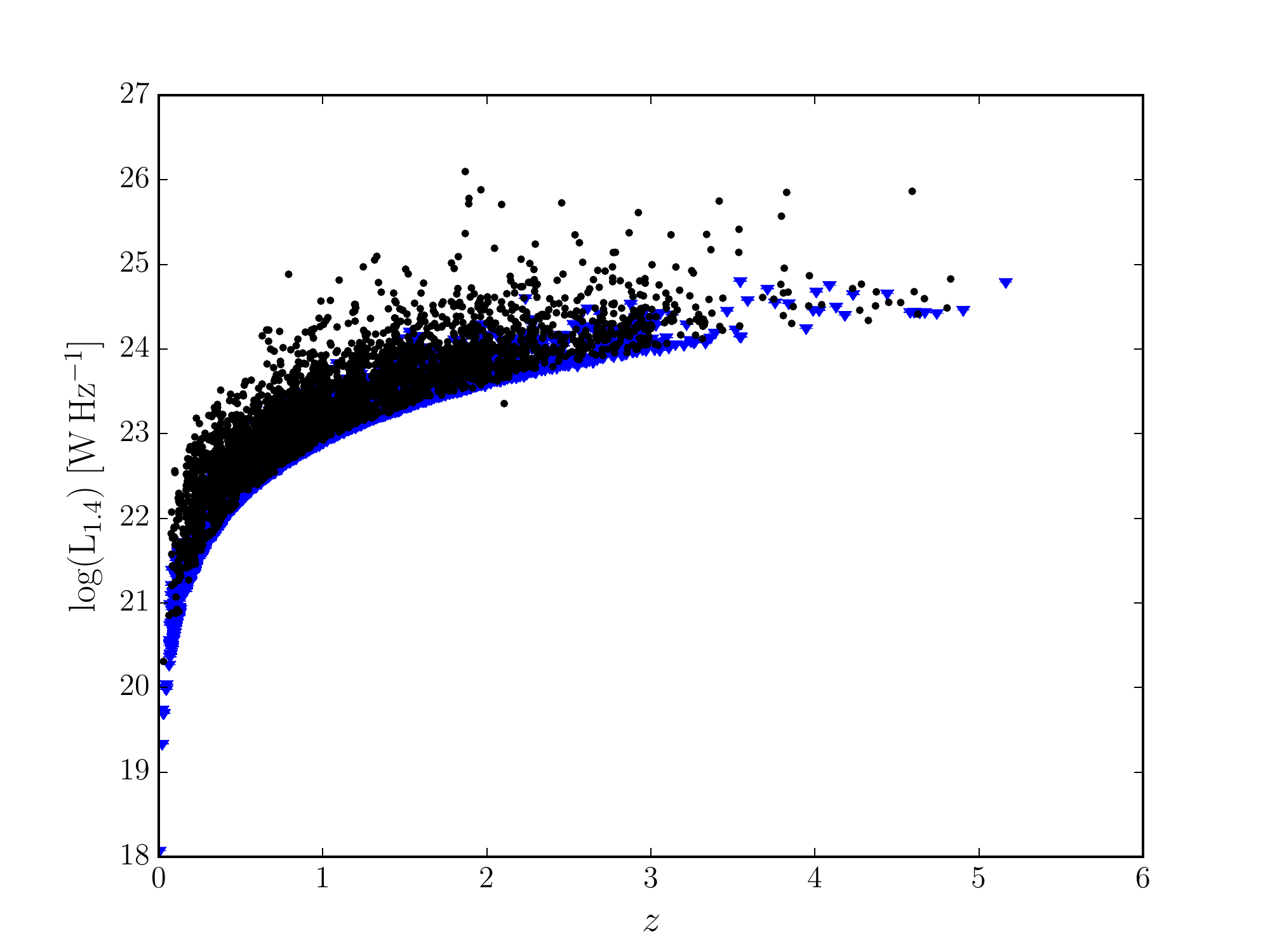}

\protect\protect\protect\protect\caption{Total infrared luminosity (left) and the 1.4\,GHz luminosity (right)
versus redshift for all star-forming objects in the jointly-selected
sample. Limits are shown as arrows for non-detections. \label{fig:lum_vs_z}}
\end{figure*}

Figure \ref{fig:Radio-vs-infrared-lum} shows the $L_{{\rm TIR}}$
versus \lrad, including limits, for the star-forming sources in the
jointly-selected sample. The $L_{{\rm TIR}}$ and \lrad\ versus
redshift are shown in Figure \ref{fig:lum_vs_z}.

We have verified that the particular choice of template suite used
for SED fitting does not have a significant impact upon the derived
infrared luminosities. For a random subsample of 100 objects, we have
recomputed the \ltir\ by using the SED template library of \citet{dale02}.
We find good agreement with the \textsc{magphys}-derived \ltir, with
a median difference of 0.027~dex and a scatter of 0.39~dex. Furthermore,
we verified that the \ltir\ estimates derived from \textsc{magphys}
are consistent with those calculated by using SED templates from \citet{chary01},
which rely on the 24\,$\mu$m detection as a proxy for the \ltir\
at $z<2$. We found no offset and a 1$\sigma$ dispersion of $\sim$0.3\,dex
between the two \ltir\ estimates. This agreement has also been determined
in previous papers (e.g. \citealt{berta13}; \citealp{delvecchio17}
and references therein).

Along with the total infrared luminosity (which we have defined as
8-1000\,$\mu$m), we also calculate the far-infrared (FIR) luminosity
($L_{{\rm FIR}}$) by integrating the star-forming template over 42-122\,$\mu$m
in the rest-frame. The median difference between the total and far-infrared
luminosities is 0.30~dex i.e.~$\langle\log(L_{{\rm TIR}})-\log(L_{{\rm FIR}})\rangle=0.30$.
The direct calculation of $L_{{\rm FIR}}$ allows for ease of comparison
with previous studies of the IRRC in the literature which have limited
their analyses to the FIR in order to avoid AGN contamination at the
shorter wavelengths (e.g.~\citealt{magnelli15,yun01}). We are not
inhibited by this issue due to our extensive AGN identification process
and our ability to decompose the origin of the emission using the
multi-component \textsc{sed3fit} fitting for such objects (see Section
\ref{sub:Galaxy-classification}).

\subsubsection{Survey sensitivity comparison \label{sub:Survey-sensitivity-comparsion}}

\begin{figure}
\includegraphics[scale=0.45]{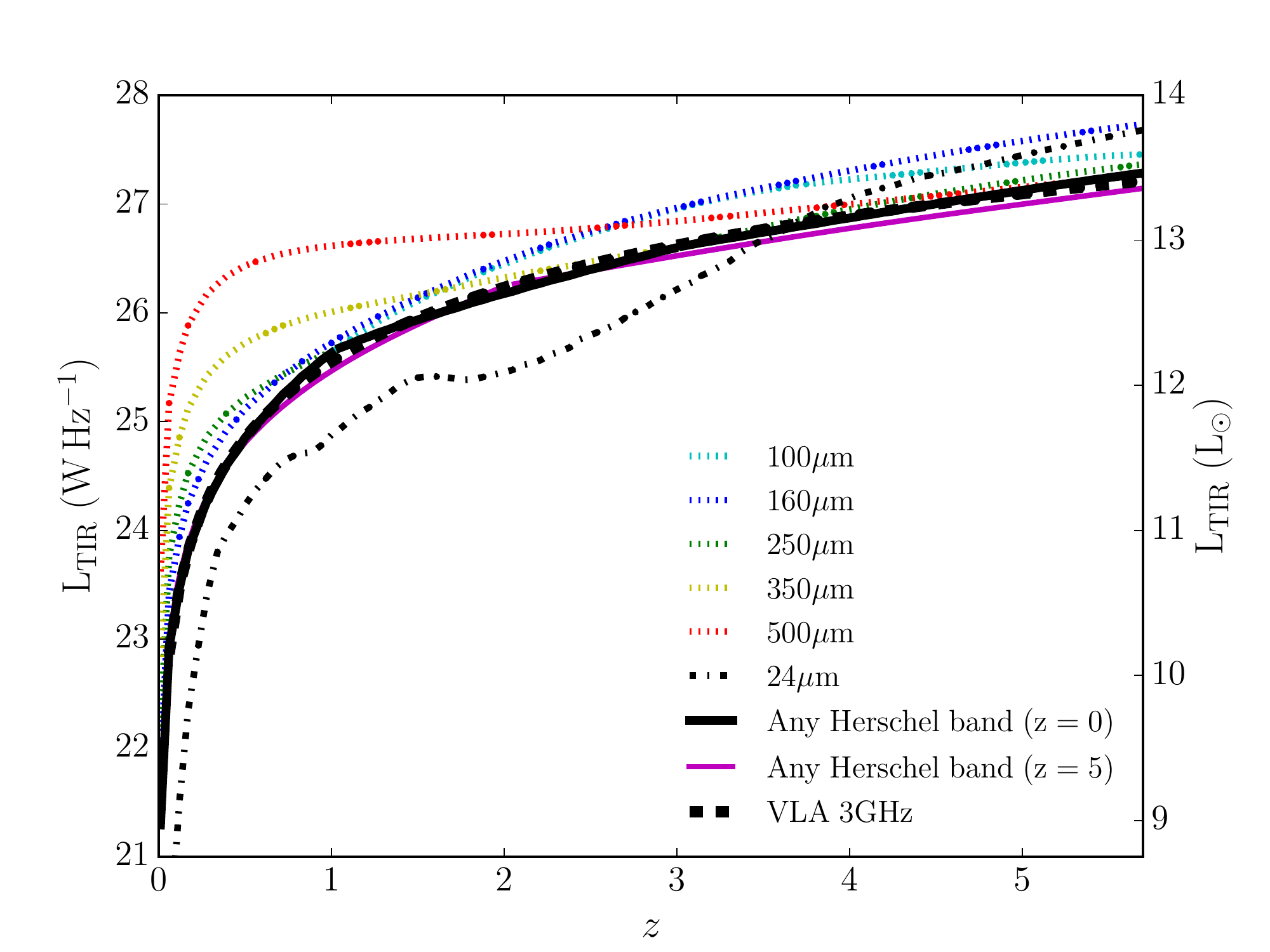}\protect\protect\protect\protect\caption{Total infrared luminosity limit of various data sets. Dashed, coloured
lines show the limit in various Herschel bands assuming a $z=0$ galaxy
template (see text, Section \ref{sub:Survey-sensitivity-comparsion},
for details). The black line traces the lowest coloured line at each
redshift and represents the sensitivity limit of the infrared-selected
sample. The magenta line is the equivalent using $z=5$ templates.
The 5\,$\sigma$ sensitivity limit of the Spitzer 24\,$\mu$m data
is shown as the black dot-dashed line. The sensitivity limit of the
VLA 3\,GHz Large Project (dashed black line) is also shown, assuming
$q_{{\rm TIR}}=2.64$ \citep{bell03} and a radio spectral index of
$\alpha=-0.7$. \label{fig:sensitivity_comparison}}
\end{figure}

The luminosity limits of the infrared and radio surveys are compared
in Figure \ref{fig:sensitivity_comparison}. The dashed, coloured
lines show the 5$\sigma$ detection limits in each Herschel band.
These have been calculated assuming a `typical' $z=0$ galaxy template
found by averaging the models of \citet{bethermin13}%
\footnote{Galaxy templates by \citet{bethermin13} at $0\leq z\leq5$ are publicly
available at ftp://cdsarc.u-strasbg.fr/pub/cats/J/A\%2BA/557/A66/%
} for normal star-forming objects on the galaxy main sequence. The
solid black line traces the lowest dashed, coloured line at each redshift.
For comparison, the equivalent line assuming $z=5$ templates is also
shown but does not differ significantly to the $z=0$ case. This represents
the lower limit for a galaxy to enter our infrared-selected sample
as it must be detected at $\geq5\sigma$ in at least one Herschel
band. However, we stress that this serves only as a rough guide for
comparison. In reality, different best-fitting galaxy templates apply
to different sources, meaning that it is possible for the $L_{{\rm TIR}}$
of a particular object to be significantly lower than the predicted
limit, while still being present in our infrared-selected sample.

The dashed black line in Figure \ref{fig:sensitivity_comparison}
shows the sensitivity of the 3\,GHz data, assuming a spectral index
of $\alpha=-0.7$ and a local conversion factor of $q_{{\rm TIR}}=2.64$
(\citealt{bell03}; see Section \ref{sub:qtir} below for the definition
of $q_{{\rm TIR}}$). We see that the sensitivities of the 3\,GHz
and Herschel surveys are well-matched out to high redshift. However,
the 24$\mu$m data, which have been used as a prior catalogue for
the infrared-selected sample, are more sensitive than both the radio
and Herschel data. In fact, 85\% of star-forming galaxies in the radio-selected
sample are detected in this 24\,$\mu$m data. Thus, most radio-detected
objects are in fact detected to some extent in the infrared, as expected.

\section{Results}

\subsection{IR-radio correlation redshift trends\label{sub:qtir} }

\begin{figure*}
\begin{centering}
\includegraphics[bb=70bp 80bp 1400bp 950bp,clip,scale=0.34]{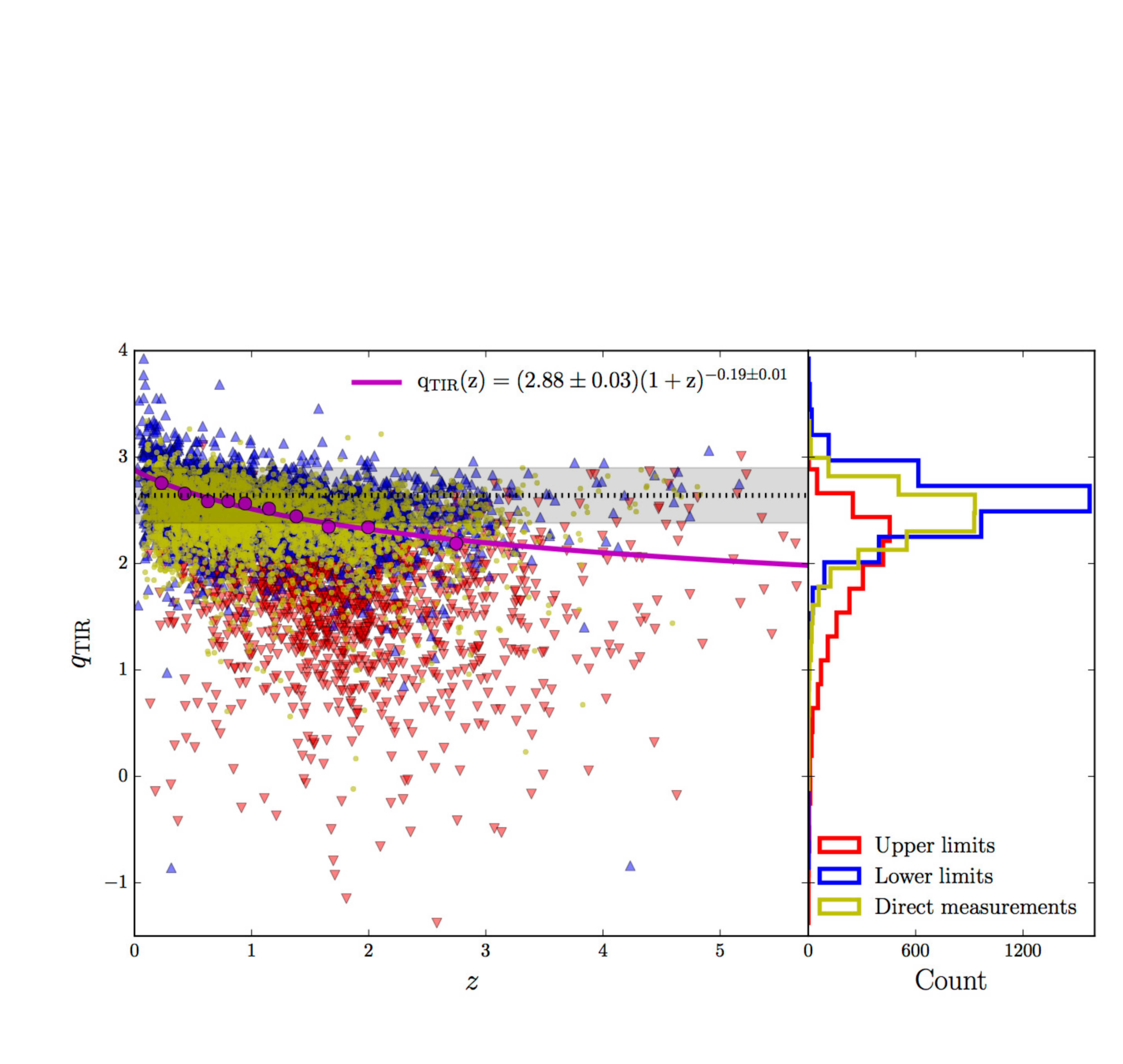} 
\par\end{centering}

\protect\protect\protect\protect\caption{IRRC (\qtir) versus redshift for star-forming galaxies. Objects with
detections in both the infrared and radio have directly-constrained
values of \qtir\ and are shown as yellow points. Objects only detected
in the radio are upper limits and shown as red triangles. Objects
only detected in the infrared are lower limits and shown as blue triangles.
A doubly-censored survival analysis has been used to calculate the
median \qtir\ within redshift bins, indicated by the magenta points.
Error bars (smaller than the magenta points here) represent the 1\,$\sigma$
dispersion calculated via the bootstrap method. The magenta line shows
the power-law fit to these. The black dotted line and grey shaded
area are the local value of \citet{bell03} (\qtir($z\thickapprox0$)$=2.64\pm0.02$)
and associated spread (0.26), respectively. In the right-hand panel,
the \qtir\ distribution is shown separately for direct measurements,
upper limits and lower limits.\label{fig:qz-main}}
\end{figure*}

The IRRC can be quantified by the parameter $q_{{\rm TIR}}$, defined
as the logarithmic ratio of the total infrared ($8-1000$\,$\mu$m)
and 1.4\,GHz luminosities:

\begin{equation}
q_{{\rm TIR}}=\log(\frac{L_{{\rm TIR}}}{3.75\times10^{12}\,{\rm Hz}})-\log(\frac{L_{1.4{\rm GHz}}}{{\rm W\, Hz}^{-1}}).\label{eq:q}
\end{equation}

We note that the $L_{{\rm TIR}}$ (in unit W) are divided by the central
frequency of $3.75\times10^{12}$\,Hz such that $q_{{\rm TIR}}$
becomes dimensionless.

\begin{figure}
\includegraphics[scale=0.45]{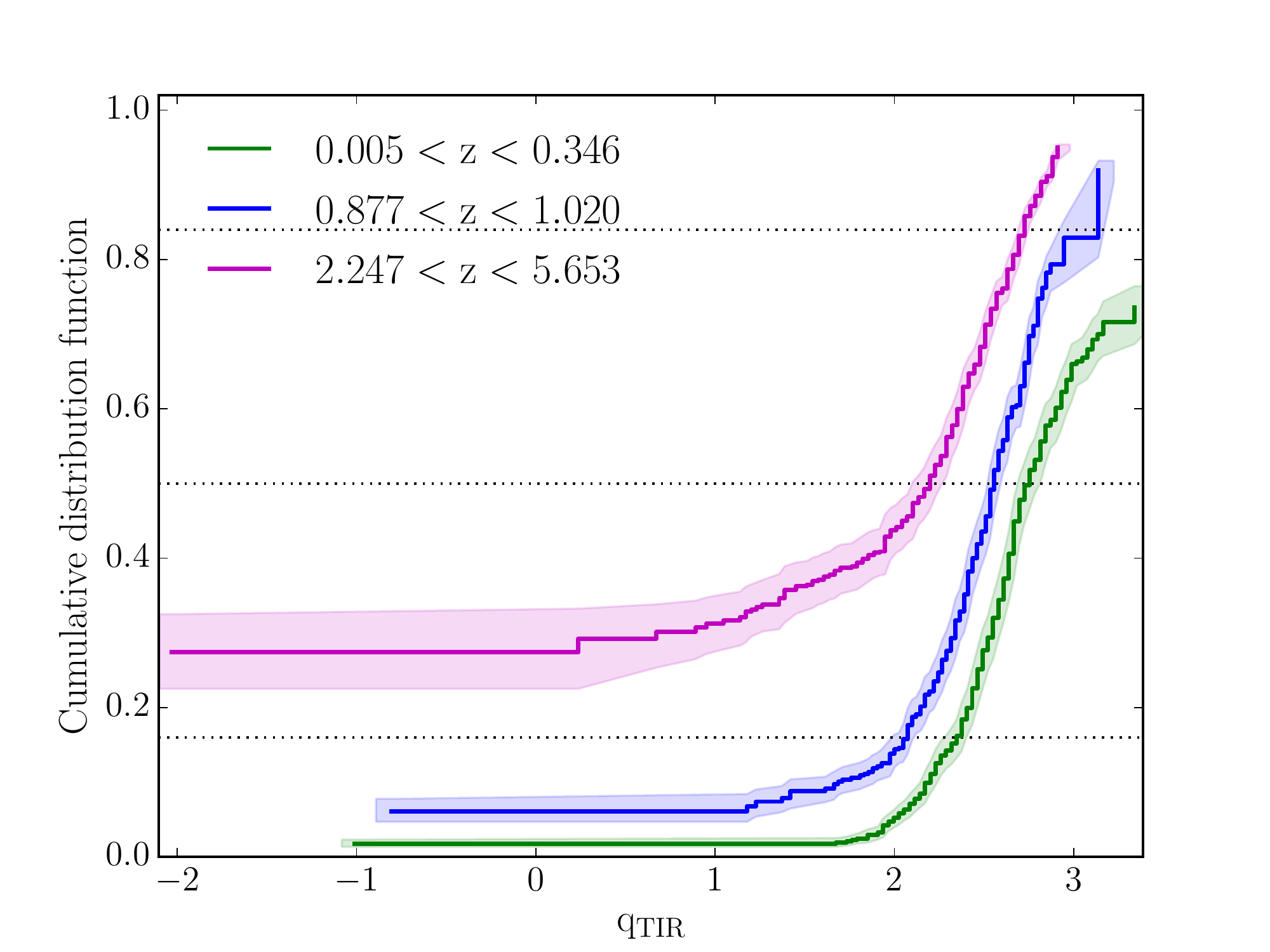}

\protect\protect\protect\protect\caption{Cumulative distribution functions produced via the doubly-censored
survival analysis within the first, fifth and tenth redshift bins.
The plots show the fraction of data with \qtir\ values less than
the value indicated on the lower axis. Shaded regions indicate the
95\% confidence interval. The 16th, 50th and 84th percentiles are
indicated by the bottom, middle and top dotted lines, respectively.\label{fig:cdf}}
\end{figure}

Figure \ref{fig:qz-main} shows the \qtir\ of all 9,575 star-forming
galaxies in our jointly-selected sample, as a function of redshift.
We have a well-populated sample out to $z\sim3$, with direct detections
in both the infrared and radio data. Upper and lower limits on \qtir\ are
also indicated in the plot. We split the data into ten redshift bins
such that they contain equal numbers of galaxies. To incorporate the
lower and upper limits when calculating the median \qtir\ in each
bin, we have employed a doubly-censored survival analysis, as presented
in \citet{sargent10a}. The basic principle of this method is that
the code (written in Perl/PDL by MTS) redistributes the limits, assuming
they follow the underlying distribution of the directly-constrained
values. This results in a doubly-censored distribution function, as
described in Schmitt (1985). An example of the cumulative distribution
function and associated 95\% confidence interval determined for several
redshift bins are shown in Figure \ref{fig:cdf}.

We use a bootstrap approach to estimate uncertainties on \qtir\ in
each redshift bin by repeating the survival analysis 100 times. In
each instance, the values of \ltir\ are randomly sampled from a
Gaussian distribution with a mean equal to the directly-constrained
nominal value and a dispersion equal to the measurement error on the
nominal value. The $S_{3{\rm GHz}}$ measurements are also sampled
in the same manner, and the flux limits are again sampled from the
distribution of optimal mosaic resolutions (see Section \ref{sub:3GHz-lims}
and Figure \ref{fig:convol-cdf}). These values are then used for
the calculation of the \qtir\ measurements or limits and the doubly-censored
survival function is regenerated.

\begin{figure}
\includegraphics[scale=0.45]{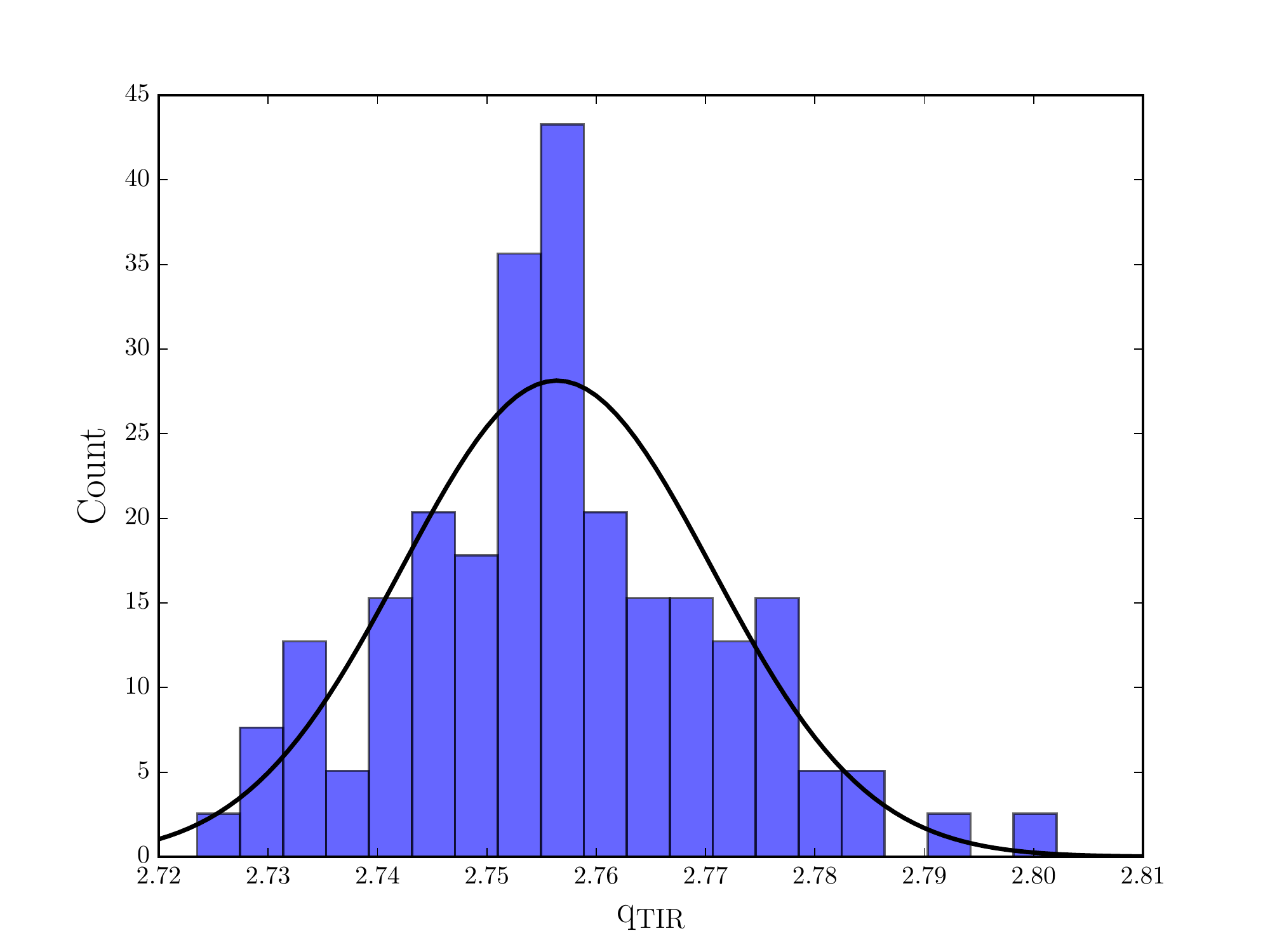}

\protect\protect\protect\caption{Distribution of the median statistic of the doubly-censored survival
function generated by resampling \qtir\ 100 times. This particular
distribution is for the $0.005<z<0.346$ redshift bin. A Gaussian
function is fit to the distribution and used to determine the final
average value of \qtir\ and its uncertainty. \label{fig:q100}}
\end{figure}

The median statistic in a given instance is the value of the 50th
percentile of the survival distribution of \qtir\ (middle dotted
line in Figure \ref{fig:cdf}). Figure \ref{fig:q100} shows an example
of the resultant distribution of the 100 median \qtir\ measurements
in a particular redshift bin and a Gaussian fit to this distribution.
The mean of this Gaussian fit provides the final average \qtir\ measurement
within the redshift bin. The 1$\sigma$ dispersion of the Gaussian
($\sim$0.01 on average) is combined in quadrature with the statistical
error on the median output from the survival analysis (indicated by
the shaded regions in Figure \ref{fig:cdf}; $\sim$0.05 on average)
to give the final uncertainty on the average \qtir. These average
values and uncertainties are reported in Table \ref{tab:qtable} and
shown in Figure \ref{fig:qz-main}. The 16th and 84th percentiles
of the survival function are also quoted, as well as the spread of
\qtir\ (i.e.~$P_{84}-P_{16}$) in each bin. We note that the survival
analysis does not constrain some of these parameters in some redshift
bins, due to the number and distribution of the limits in that bin.

We fit a power-law function to the average values of \qtir, weighting
by the uncertainty, and find a small but statistically-significant
variation of \qtir\ with redshift: $q_{{\rm TIR}}(z)=(2.88\pm0.03)(1+z)^{-0.19\pm0.01}$.
The errors here are the 1$\sigma$ uncertainty from the power-law
fit.

\begin{table*}
\protect\caption{Median value of $z$ and \qtir\ and number of star-forming galaxies
in each redshift bin.}

\begin{tabular}{cccccccccc}
\hline 
$z$  & median($z$)  & $N_{{\rm total}}$  & $N_{{\rm meas}}$  & $N_{{\rm upper}}$  & $N_{{\rm lower}}$  & $q_{{\rm TIR}}$  & $P_{16}$  & $P_{84}$  & $(P_{84}-P_{16})$\tabularnewline
\hline 
0.005-0.346  & 0.23  & 958  & 425 (44\%)  & 63 (7\%)  & 470 (49\%)  & $2.76_{-0.06}^{+0.07}$  & 2.35  & $>3.34$  & $>0.99$\tabularnewline
0.346-0.527  & 0.428  & 957  & 421 (44\%)  & 117 (12\%)  & 419 (44\%)  & $2.66_{-0.05}^{+0.03}$  & 2.28  & 3.27  & 0.99\tabularnewline
0.527-0.704  & 0.626  & 957  & 402 (42\%)  & 153 (16\%)  & 402 (42\%)  & $2.59_{-0.06}^{+0.03}$  & 2.17  & 2.94  & 0.77\tabularnewline
0.704-0.877  & 0.801  & 958  & 376 (39\%)  & 163 (17\%)  & 419 (44\%)  & $2.59_{-0.05}^{+0.03}$  & 2.09  & 2.92  & 0.83\tabularnewline
0.877-1.020  & 0.944  & 957  & 361 (38\%)  & 172 (18\%)  & 424 (44\%)  & $2.56_{-0.04}^{+0.04}$  & 2.07  & 3.14  & 1.07\tabularnewline
1.020-1.245  & 1.149  & 958  & 312 (33\%)  & 239 (25\%)  & 407 (42\%)  & $2.52_{-0.06}^{+0.03}$  & 1.45  & 2.93  & 1.48\tabularnewline
1.245-1.509  & 1.381  & 957  & 299 (31\%)  & 284 (30\%)  & 374 (39\%)  & $2.44_{-0.06}^{+0.04}$  & $<-0.54$  & 2.84  & $>3.38$\tabularnewline
1.509-1.835  & 1.657  & 958  & 295 (31\%)  & 322 (34\%)  & 341 (36\%)  & $2.35_{-0.05}^{+0.08}$  & $<-1.83$  & 2.81  & $>4.64$\tabularnewline
1.835-2.247  & 1.995  & 957  & 355 (37\%)  & 282 (29\%)  & 320 (33\%)  & $2.34_{-0.05}^{+0.06}$  & 0.18  & 2.73  & 2.55\tabularnewline
2.247-5.653  & 2.746  & 958  & 336 (35\%)  & 380 (40\%)  & 242 (25\%)  & $2.19_{-0.07}^{+0.10}$  & $<-2.03$  & 2.73  & $>4.76$\tabularnewline
\hline 
\end{tabular}\centering{} \tablefoot{A break-down of the number of sources
into those with directly measured \qtir\ values ($N_{{\rm meas}}$),
upper limits on \qtir\ ($N_{{\rm upper}}$) or lower limits on \qtir\ ($N_{{\rm lower}}$)
is shown, with the fraction of the total shown in brackets. The \qtir\
is calculated using a doubly-censored survival analysis to incorporate
lower and upper limits. Uncertainties on \qtir\ are calculated using
a bootstrap approach and incorporate statistical, measurement and
systematic errors. A radio spectral index of $\alpha=-0.7$ has been
assumed where it is unknown. The 16th and 84th percentiles ($P_{16}$
and $P_{84}$) on the measurement of \qtir\ in each redshift bin
are given, as determined via the cumulative distribution function
output by the survival analysis. $(P_{84}-P_{16})$ is quoted to indicate
the spread of the population. We note that limits are given when a
value is not constrained by the survival analysis.\label{tab:qtable}} 
\end{table*}

\section{Discussion\label{sec:Discussion}}

\subsection{Comparison with previous studies\label{sub:lit-comparison}}

\begin{figure}
\begin{raggedright} \includegraphics[scale=0.45]{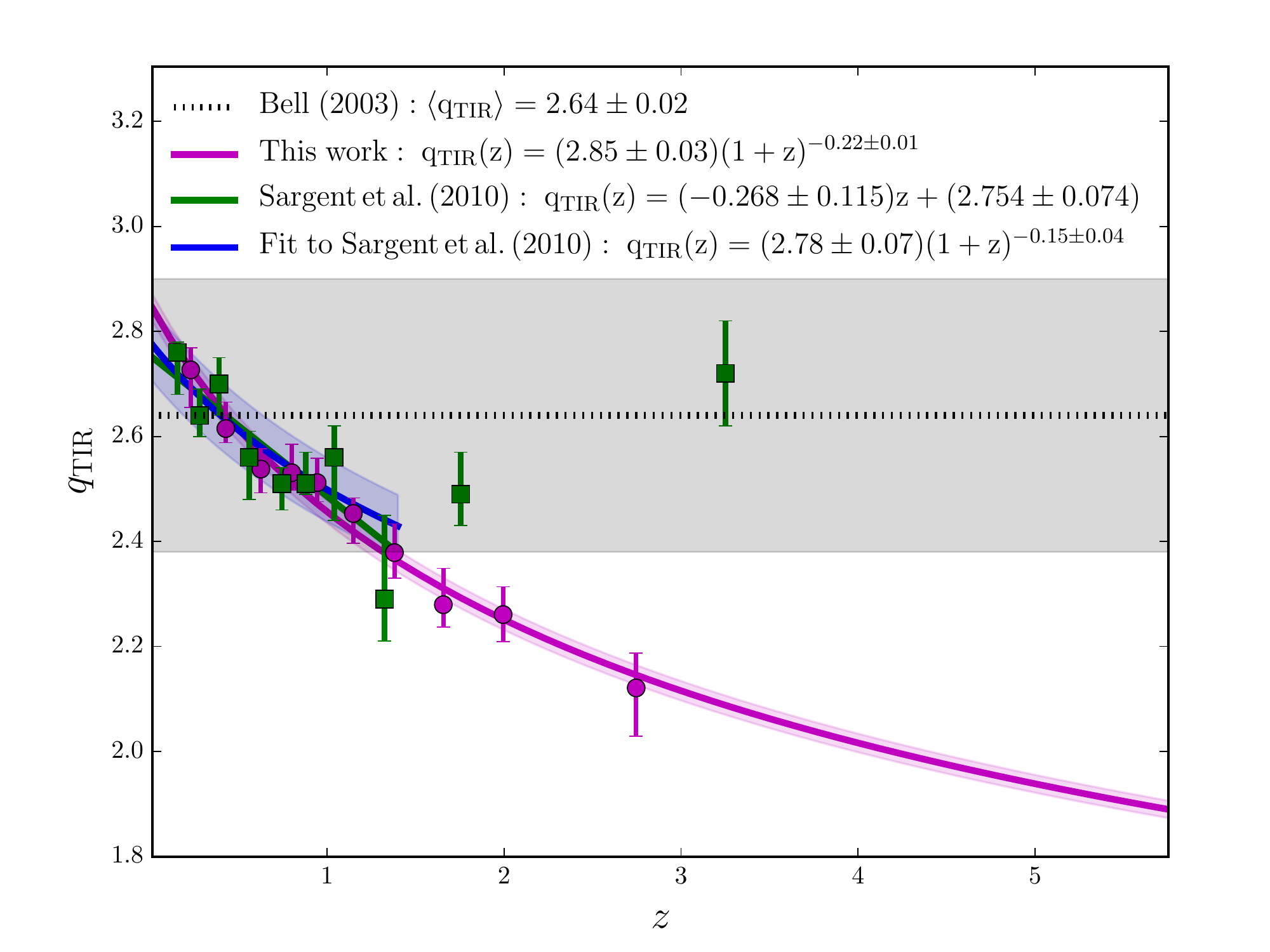}

\end{raggedright}

\protect\protect\protect\protect\caption{Evolution of \qtir\ in comparison with the results of \citet{sargent10a}.
The magenta points and fit show the results from this work using a
full survival analysis, as in Figure \ref{fig:qz-main}, however a
spectral index of $\alpha=-0.8$ has now been assumed for objects
not detected at 1.4\,GHz. The measurements of \citet{sargent10a}
and their linear fit are shown by the green points and line. A power-law
evolution to the individual measurements of \citet{sargent10a} is
shown by the blue line, for ease of comparison. The shaded magenta
and blue regions show the 1$\sigma$ uncertainty regions calculated
by propagating the errors on the corresponding fitting parameters.
The local measurement and spread (grey shading) of \citet{bell03}
are also shown.\label{fig:TIR-comparison}}
\end{figure}

\begin{figure}
\includegraphics[scale=0.45]{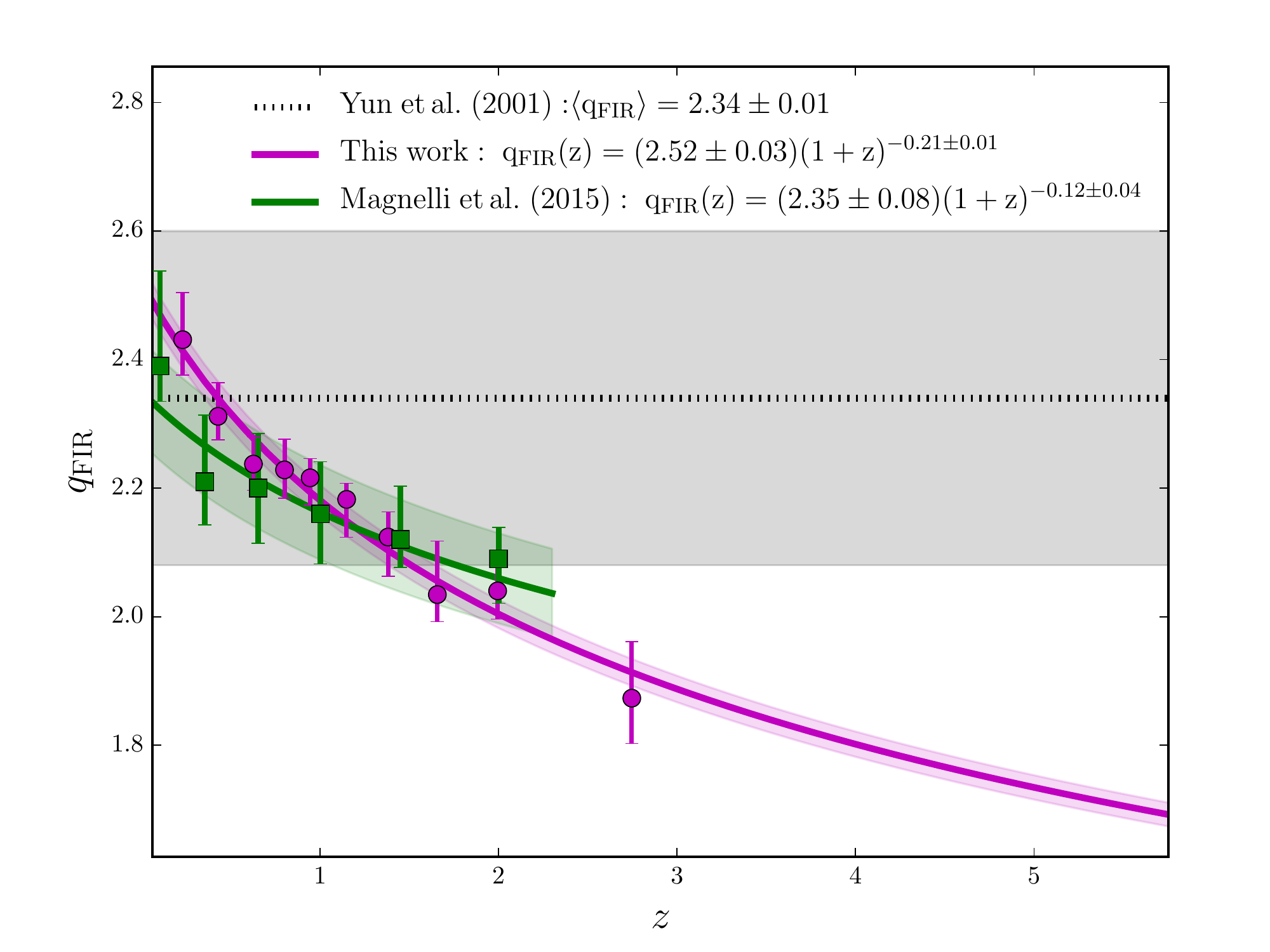}\protect\protect\protect\protect\caption{FIR-radio correlation ($q_{{\rm FIR}}$) versus redshift for star-forming
galaxies. The evolving fit generated via a survival analysis in this
work, assuming $\alpha=-0.8$ for objects not detected at 1.4\,GHz,
is shown by the magenta line. The evolution found by \citet{magnelli15}
using a stacking analysis is shown by the green points and curve.
The shaded magenta and green regions show the 1$\sigma$ uncertainty.
The local value of Yun et al.~(2001; $2.34\pm0.01$) and associated
spread (0.26) are shown by the dashed line and grey shaded area, respectively.\label{fig:FIR-comparison}}
\end{figure}

Here we compare our results to several other studies in the literature.
To reduce systematics introduced by converting between measurements
of \qtir\ and the FIR-radio correlation ($q_{{\rm FIR}}$), we have
compared our results separately to those quoted using TIR and those
using FIR. As described in Section \ref{sub:Infrared-luminosities},
we are able to directly measure the \ltir\ and $L_{{\rm FIR}}$
as a result of the SED fitting process, and therefore can directly
calculate both \qtir\ and $q_{{\rm FIR}}$. For ease of comparison,
we have also assumed a spectral index of $\alpha=-0.8$ for non-detections
at 1.4\,GHz when calculating $L_{1.4}$, as was assumed in \citet{sargent10a}
and \citet{magnelli15}. Artificial discrepancies could be introduced
if different studies assumed different spectral indices, as will be
demonstrated in Section \ref{sub:discussion-Lrad}.

As shown in Figure \ref{fig:TIR-comparison}, our calculated median
values of \qtir\ at $z<1.4$ are consistent with those of \citet{sargent10a},
who also employ a doubly-censored survival analysis to incorporate
non-detections into their measurements. At higher redshift, the increase
of \qtir\ with redshift found by \citet{sargent10a} is not consistent
with our results; a possible reason for this discrepancy is the fact
that, as noted by \citet{sargent10a}, high-quality photometric redshifts
were not available to them over this range. \citet{sargent10a} fit
a linear relation with redshift to their data up to $z=1.4$: $q_{{\rm TIR}}(z)=(-0.268\pm0.115)z+(2.754\pm0.074)$.
For ease of comparison with our adopted functional form of the fit,
we also fit a power-law relation in $(1+z)$ to their data: $q_{{\rm TIR}}(z)=(2.78\pm0.07)(1+z)^{-0.15\pm0.04}$.
The slope of this best fit\textbf{ }is slightly flatter than, but
consistent within 2$\sigma$,\textbf{ }with our results based on a
doubly-censored survival analysis using $\alpha=-0.8$: $q_{{\rm TIR}}(z)=(2.85\pm0.03)(1+z)^{-0.22\pm0.01}$.

The redshift trend that we find is also in agreement with the recent
results of \citet{magnelli15}, as shown in Figure \ref{fig:FIR-comparison}.
These authors use a stacking analysis to examine the evolution of
the FIR-radio correlation. They find $q_{{\rm FIR}}(z)=(2.35\pm0.08)(1+z)^{-0.12\pm0.04}$.
Although our measurements within each redshift bin for star-forming
galaxies using a survival analysis are largely consistent with those
of \citet{magnelli15}, the fitted trend we derive has a slightly
higher normalisation and steeper slope: $q_{{\rm FIR}}(z)=(2.52\pm0.03)(1+z)^{-0.21\pm0.01}$.
This is within 2$\sigma$ agreement with our results. We also note
that \citet{calistro-rivera17} find a similarly decreasing trend
of \qtir$(z)$ for a radio-selected sample of star-forming galaxies
in the Bo\"otes field.

\begin{figure}
\includegraphics[scale=0.45]{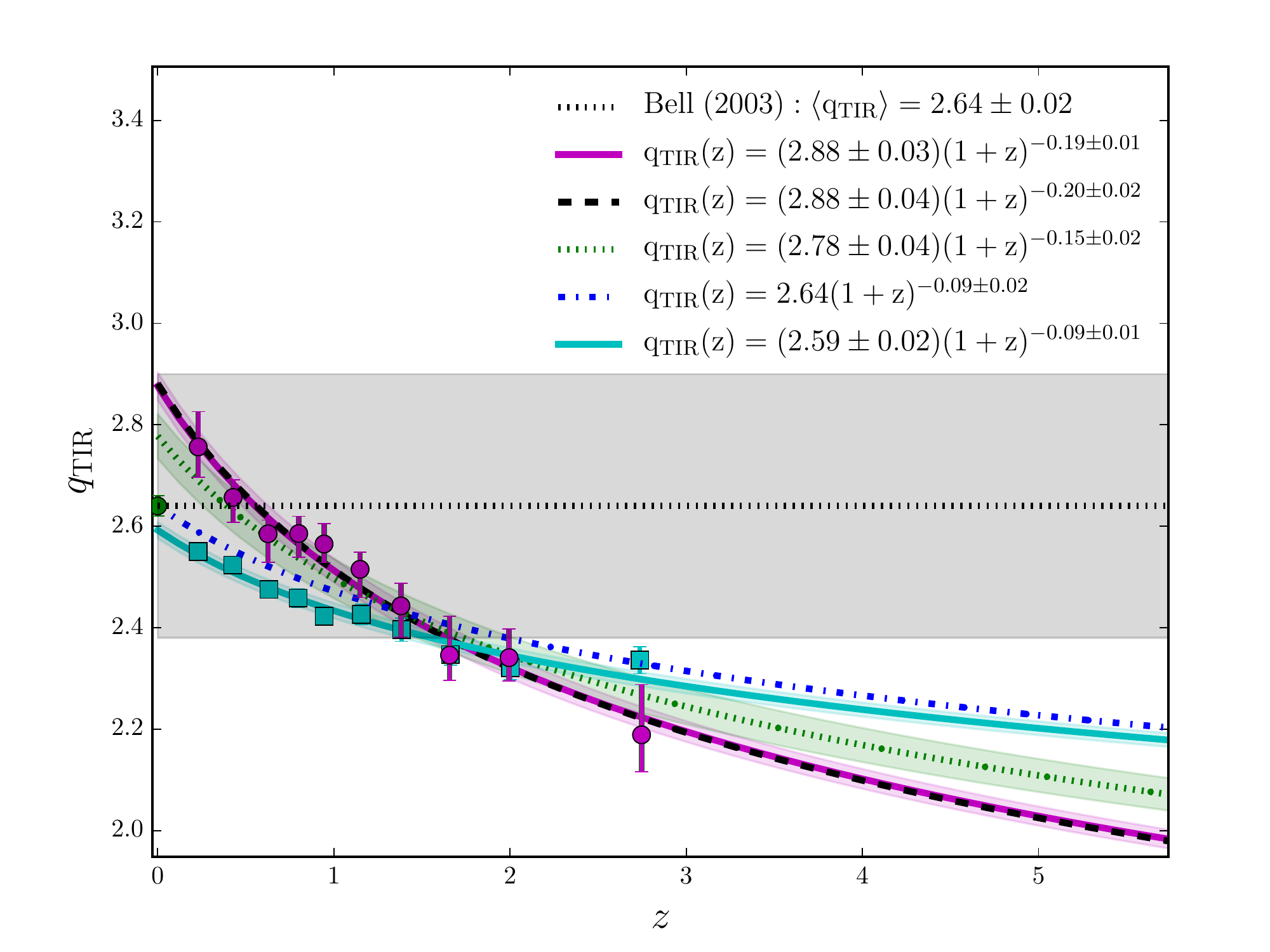}\protect\protect\protect\protect\caption{\qtir\ versus redshift for star-forming galaxies and associated
power-law fits derived using: all data points (magenta points and
solid line; 1$\sigma$ uncertainty region shaded), excluding the lowest
redshift bin (black dashed line), including the local value of \citet{bell03}
(i.e.~fitting to all the magenta points as well as the green point;
green dotted line with 1$\sigma$ uncertainty region shaded), and
anchoring to the local value of \citep{bell03} by fitting to the
function $2.64(1+z)^{x}$ where $x$ is the free parameter (blue dot-dashed
line). Also shown is the median in each redshift bin calculated using
only directly-measured values (i.e.~without applying a survival analysis;
cyan squares), and the associated fit (solid cyan line).\label{fig:qz-local-anchor}}
\end{figure}

As can be seen in Figures \ref{fig:TIR-comparison} and \ref{fig:FIR-comparison},
our measurements of \qtir\ and $q_{{\rm FIR}}$ in the lowest redshift
bin are slightly higher (by more than the 1$\sigma$ uncertainty)
than the local values of \citet{bell03} and \citet{yun01}, respectively.
While we have attempted to account for resolution bias in the radio
data, it is possible that we still miss emission from the most extended
sources, which are likely to be present at the lowest redshifts. However,
our low redshift measurements are consistent with those of \citet{sargent10a}
who used radio data at a lower resolution ($\sim1.5''$) and are therefore
less affected by resolution bias. It is therefore unlikely that our
results are significantly impacted by resolution bias. It is also
possible that our results are affected by issues related to blending
in the Herschel maps.

If we exclude the first redshift bin from the fitting procedure, we
find that the \qtir$(z)$ trend is not altered within 1$\sigma$,
as seen in Figure \ref{fig:qz-local-anchor}. To examine the effect
of including the local value in the analysis, we include a \qtir$(z=0)=2.64$
data point when performing the fit to \qtir$(z)$. As shown in Figure
\ref{fig:qz-local-anchor}, the resulting \qtir$(z)$ trend is slightly
flatter: $q_{{\rm TIR}}(z)=(2.78\pm0.04)(1+z)^{-0.15\pm0.02}$. To
examine the extreme case, we `anchor' the trend to the local value
by fitting the expression $q_{{\rm TIR}}(z)=2.64(1+z)^{x}$, where
$x$ is the free parameter. We still find a decrease in\textbf{ }\qtir\
with redshift to a 5$\sigma$ significance level. This suggests that
a decreasing trend of \qtir$(z)$ is always observed, with the exponent
of $(1+z)$ between -0.20 and -0.09, regardless of the treatment of
the low-redshift measurement.

\subsection{Impact of upper and lower limits}

In Table \ref{tab:qtable} it can be seen that the fraction of upper
and lower limits on \qtir\ in a given bin changes with redshift.
It is possible that the apparent decrease in \qtir\ with increasing
redshift could be somehow driven by the changing fraction of limits.
To examine the extreme case, we ignore all limits and calculate the
median of only directly-constrained values of \qtir\ in each redshift
bin. These values are shown in Figure \ref{fig:qz-local-anchor} with
error bars representing the standard error on the median. Using these
measurements we find a trend of $q_{{\rm TIR}}(z)=(2.59\pm0.02)(1+z)^{-0.09\pm0.01}$.
This fit is flatter than that found when non-detections are correctly
accounted for using a survival analysis, producing smaller \qtir\
values particularly at lower redshifts. This indicates that accounting
for non-detections (limits) in such an analysis has a profound impact
on the results.

It is interesting to note that the exponent of the \qtir\ trend
found when excluding limits agrees with that found in Section \ref{sub:lit-comparison}
through anchoring to the local value while incorporating limits. It
is perhaps worth noting that these studies at $z\sim0$ also dealt
only with direct detections and not with limits. Overall, our conclusion
again is that a decrease in\textbf{ }\qtir\ with redshift is always
observed, with the value of the $(1+z)$ exponent varying between
-0.20 and -0.09, depending on the particular treatment of non-detections
and low-redshift data.

We also note that our survival analysis produces results consistent
with those of \citet{magnelli15} who accounted for limits using the
independent approach of stacking. \citet{mao11} also find that the
use of a survival analysis and a stacking analysis to account for
limits in studies of \qtir$(z)$ give similar results. Of course,
the optimal solution would be to have direct detections available
for a complete sample. However, such data are not yet available. Thus,
despite our attempts to account for the non-detections through a survival
analysis, we acknowledge that our results could still be affected
by the sensitivity limitations of the data.

Related to this, we also acknowledge the strong trend between redshift
and luminosity of objects in our sample, resulting from the data sensitivity
limits. We have performed a partial correlation analysis (see e.g.
\citealt{macklin82}) to determine whether a correlation between \qtir\
and redshift exists when the dependence on radio or infrared luminosities
are removed. However, our results are inconclusive due to biases introduced
by the flux limit of our sample. Breaking this degeneracy would require
a well-populated, complete sample spanning several orders of magnitude
in both radio and infrared luminosity at each redshift. We therefore
emphasise that the results we present in this paper are based upon
the assumption of\textbf{ }a luminosity-independence of\textbf{ }\qtir\
at all redshifts.

\subsection{AGN contributions}

\subsubsection{Are many moderate-to-high radiative luminosity AGN misclassified
as star-forming galaxies?}

\begin{figure}
\includegraphics[scale=0.5]{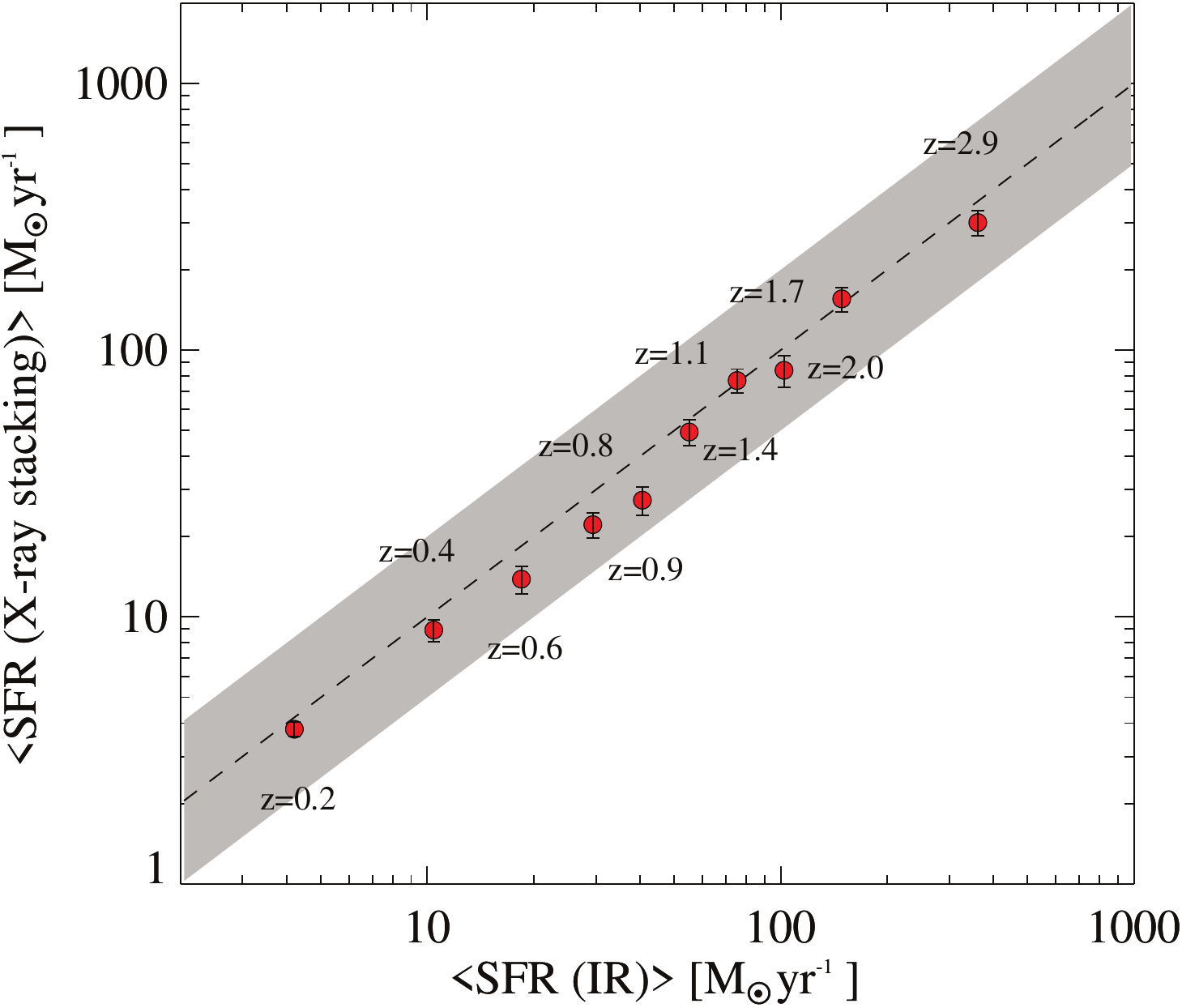}

\protect\protect\protect\protect\caption{Star formation rate predicted from the infrared emission of the Herschel-detected
star-forming galaxies in our sample, compared to that predicted via
X-ray stacking. The grey region encloses a factor of two around the
1:1 relation, and corresponds to the observed scatter of the $L_{X}$-SFR
relation presented by \citet{symeonidis14}. No excess is seen in
the X-rays, indicating no appreciable contribution from AGN.\textbf{\label{fig:xstack}}}
\end{figure}

We wish to determine the extent to which AGN contamination could be
influencing our results. Although we have used all information at
hand to identify objects that are very likely to host AGN, it is still
possible that some sources in our star-forming sample have been misclassified
or contain low levels of AGN activity. We can investigate the extent
to which our sample is contaminated by misclassified HLAGN via X-ray
stacking. If misclassified AGN are present, the stacked X-ray flux
of the full sample should exceed that expected purely from star formation
processes. To test this, we used the publicly-available CSTACK%
\footnote{CSTACK was created by Takamitsu Miyaji and is available at http://lambic.astrosen.unam.mx/cstack/%
} tool to stack Chandra soft ({[}0.5-2{]}keV) and hard band ({[}2-8{]}keV)
X-ray images of all objects within each redshift bin. The stacked
count rate is converted into a stacked X-ray luminosity by assuming
a power law spectrum with a slope of 1.4, consistent with the X-ray
background (e.g.~\citealt{gilli07}). We then apply the conversion
between X-ray luminosity and SFR {derived by \citet{symeonidis14}.
This conversion was calibrated on Herschel galaxies, both detected
and undetected in X-ray, for a better characterisation of the average
L$_{X}$-SFR correlation in inactive star-forming galaxies}%
\footnote{We note that we have scaled the relation of \citet{symeonidis14}
to match the X-ray bands and spectral slope chosen here.%
}{. Figure \ref{fig:xstack} shows the SFR derived from X-ray stacking
compared to the SFR derived from infrared luminosities. The latter
was found using the conversion of \citet{kennicutt98} assuming a
\citet{chabrier03} IMF and is not expected to be significantly affected
by AGN activity and therefore solely attributable to star formation.
We find no excess in the X-ray-derived SFR with respect to the IR-derived
SFR, indicating that there are very few misclassified HLAGN in our
star-forming sample of galaxies.}

\subsubsection{Infrared-radio correlation of AGN}

\begin{figure}
\includegraphics[scale=0.45]{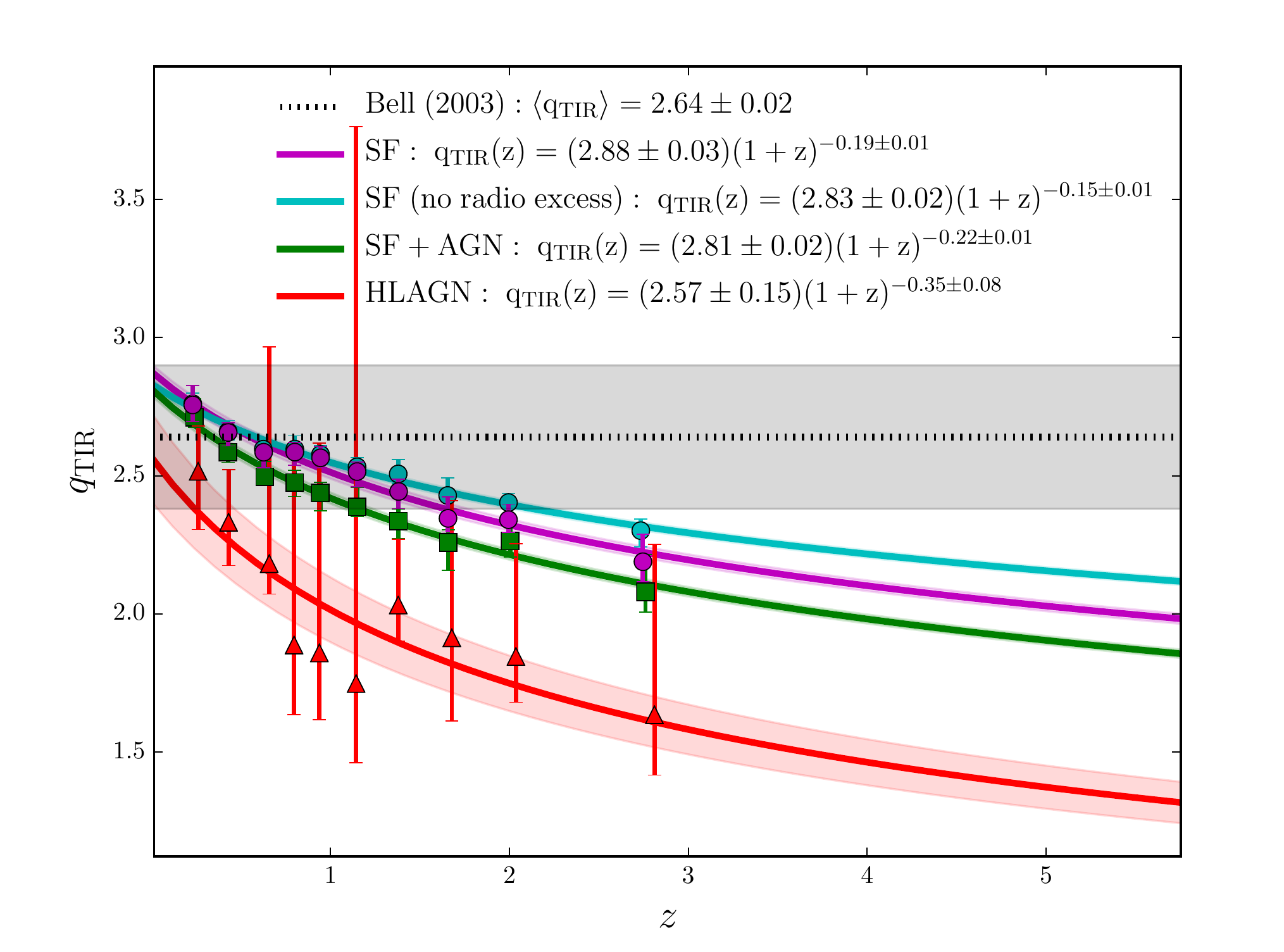}

\protect\protect\protect\protect\caption{Evolution of the IRRC for different source populations. The magenta
curve (and points) is the power-law relation found for star-forming
galaxies only, while the green curve (and squares) is that found when
AGN are included (i.e.~star-forming galaxies plus all AGN). The red
curve (and triangles) is found when only HLAGN are considered. The
cyan curve (and points) is found for the star-forming population of
galaxies, excluding those with radio excess. See text (Section \ref{sub:radio-excess})
for the definition of radio excess. Shading shows the 1$\sigma$ uncertainty
regions. \label{fig:q_v_z+agn}}
\end{figure}

\begin{figure}
\includegraphics[bb=0bp 0bp 400bp 650bp,clip,scale=0.6]{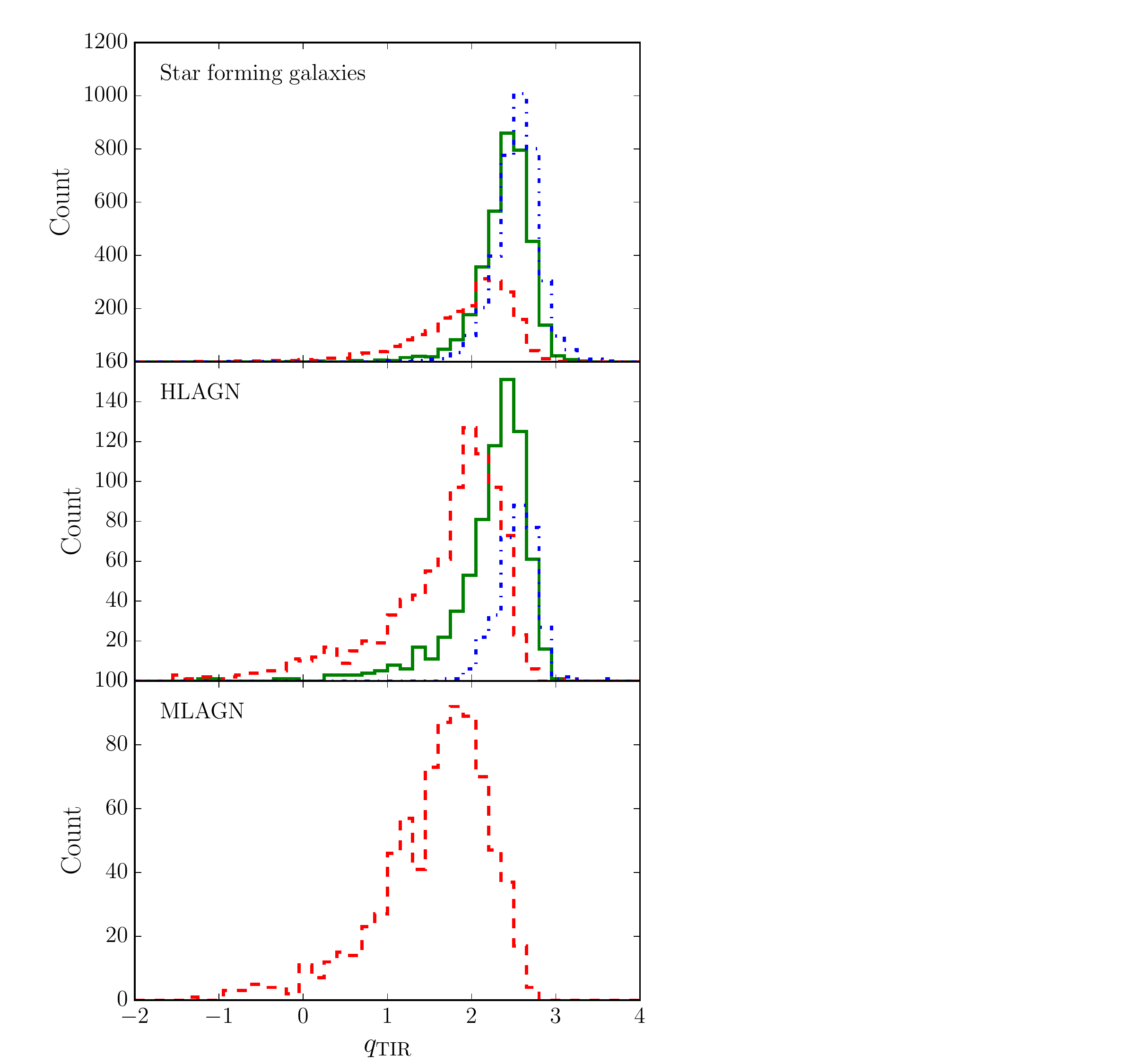}

\protect\protect\protect\caption{Distribution of direct \qtir\ measurements (solid green line), lower
limits (blue dot-dashed line) upper limits (red dashed line) shown
separately for the populations of star-forming galaxies, HLAGN and
MLAGN, as indicated.\label{fig:qdist_sf+agn}}
\end{figure}

Despite the fact that we expect minimal numbers of misclassified HLAGN,
we nonetheless investigate how the emission arising from AGN activity,
rather than star-formation processes, could impact the results. Figure
\ref{fig:q_v_z+agn} shows the resulting \qtir\ as a function of
redshift if we apply the survival analysis, described in Section \ref{sub:qtir},
to all objects in the jointly-selected sample. That is, to all star-forming
galaxies as well as all HLAGN and MLAGN (see Section \ref{sub:Galaxy-classification}).
We find only a slight ($<2\sigma$) decrease in the normalisation
of the power law fit and steepening of the slope when compared to
star-forming galaxies only. This indicates that the inclusion or exclusion
of known AGN (which only consitute 22\% of the full sample) does not
significantly impact the overall \qtir$(z)$ trend found.

If we consider only objects in the HLAGN category, the inferred trend
of \qtir\ with redshift for this population appears significantly
steeper than that for star-forming galaxies only, although is affected
by large uncertainties at higher redshifts. Overall, this suggests
that the dependence with redshift of the IRRC of HLAGN is different
to that of star-forming galaxies. We note that for this analysis,
the \ltir\ of HLAGN has been calculated by integrating only the
star-forming galaxy component of the multi-component SED template
fit determined by \textsc{sed3fit. }That is, we exclude the AGN component
and its contribution to the \ltir. See Section \ref{sub:Galaxy-classification}
and \citet{delvecchio17} for further details.

We note that, by definition, only upper limits on \qtir\ are available
for the MLAGN (see Section \ref{sub:Galaxy-classification}) and therefore
we cannot directly investigate the behaviour of this population alone.

Figure \ref{fig:qdist_sf+agn} shows the distributions of direct \qtir\
measurements and limits separately for the star-forming galaxies and
the two classes of AGN. Although the two classes of AGN comprise only
22\% of the full sample, they are responsible for many of the extreme
measurements (or limits) of \qtir. In particular, the upper limits
of the MLAGN largely sit towards lower \qtir\ values (i.e.~have
radio-excess) with respect to the \qtir\ distribution of star-forming
galaxies. The lower median \qtir, and large fraction of upper limits,
of AGN may be explained by the presence of significant AGN contribution
to the radio continuum, with a potentially lower fractional contribution
in the infrared. In particular, the far-infrared Herschel bands should
be relatively free of AGN contamination, as the thermal emission from
the dusty torus peaks in the mid-IR (e.g.~\citealt{dicken09,hardcastle09}).
Furthermore, we find no obvious bias in the directly-detected \ltir\
distribution of the AGN compared to the star-forming population. We
again note that any AGN contribution to the \ltir\ should have been
excluded via the SED-fitting decomposition mentioned above. It is
therefore possible that AGN contamination only in the radio regime
could be contributing to the observed decrease of \qtir\ with redshift.

\subsubsection{Radio-excess objects\label{sub:radio-excess}}

\begin{figure}
\includegraphics[scale=0.45]{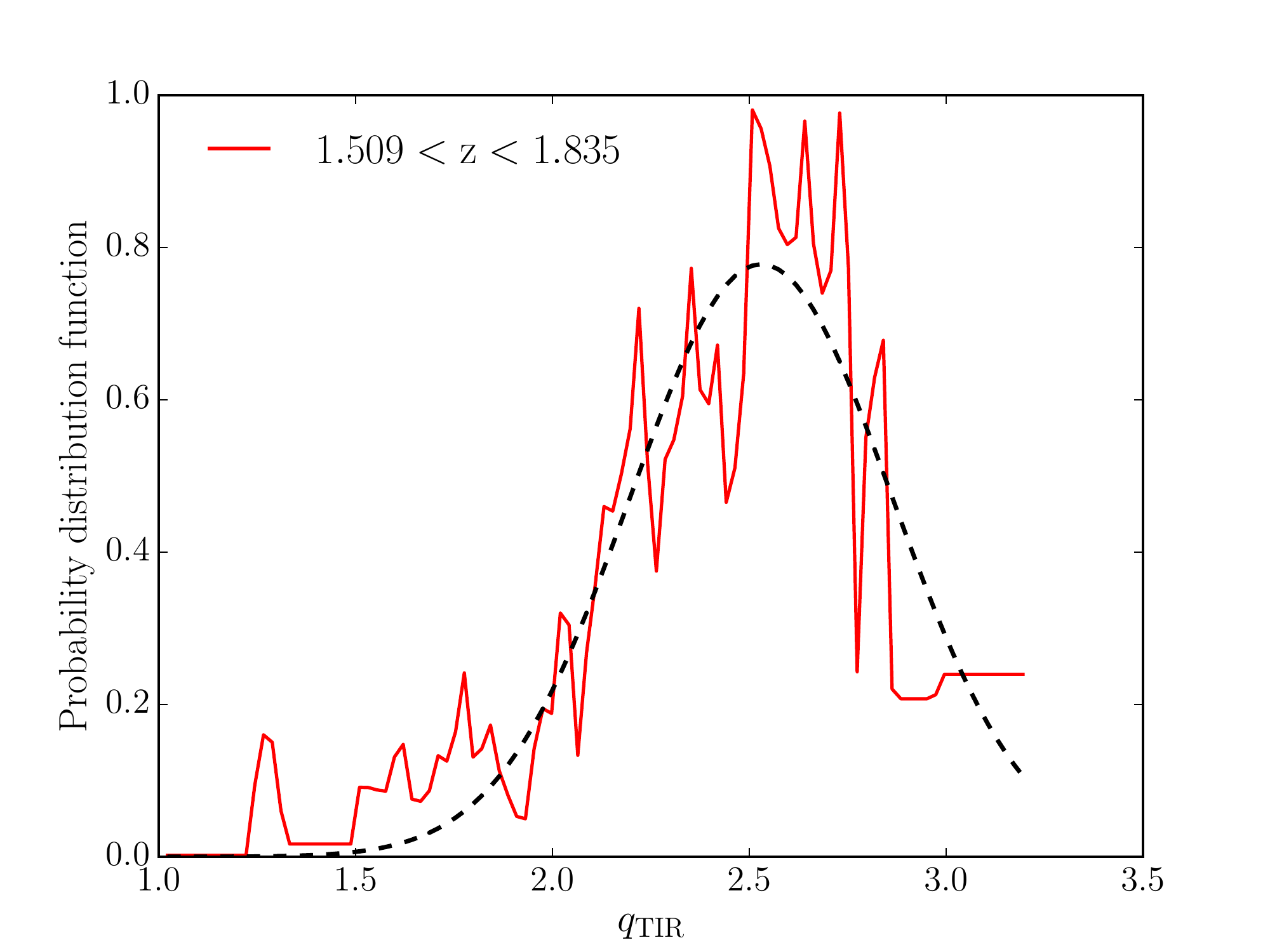}

\protect\protect\protect\caption{Probability distribution in a given redshift bin used to identify
objects with radio-excess. The probability distribution function (red
line) is generated by taking the derivative of the survival function
(a cumulative distribution) in a given redshift bin and is fitted
with a Gaussian function (black dashed line). \label{fig:sa+gauss}}
\end{figure}

It is notoriously difficult to separate AGN and star-formation contributions
to the radio when no AGN identifiers are available at other wavelengths.
Although we have identified MLAGN based upon their red optical colours
and lack of Herschel detections (see Section \ref{sub:Galaxy-classification}),
it is still possible that some objects in our star-forming galaxy
sample may also contain MLAGN which contribute only in the radio.
Such objects may be expected to show radio excess in their \qtir\
values. We therefore again examine the trend of \qtir\ versus redshift
for the star-forming population of galaxies, this time excluding objects
displaying a radio excess.\textbf{ }We define an appropriate cut to
exclude such objects in each redshift bin as follows: We take the
derivative of the survival function and then fit a Gaussian profile
to the resulting probability distribution function. An example of
this is shown in Figure \ref{fig:sa+gauss}. The dispersion ($\sigma$)
and the mean ($\mu$) of this Gaussian function are used to define
radio-excess objects as those with \qtir$<(\mu-3\sigma)$. The median
value of $\sigma$ across the redshift bins is 0.34. We then rerun
the survival analysis excluding these 510 radio excess objects (5\%
of the star-forming sample). The result, as seen in Figure \ref{fig:q_v_z+agn},
is inconsistent with the inclusion of these objects (i.e.~the full
star-forming sample), having a shallower slope: $q_{{\rm TIR}}(z)=(2.83\pm0.02)(1+z)^{-0.15\pm0.01}$.
Thus, sources with appreciable radio excess may play a role in the
observed $q_{{\rm TIR}}(z)$ trend of the star-forming sample. It
is also possible that an appreciable fraction of objects in this star-forming
sample are in fact composite systems containing (currently unidentified)
MLAGN which contribute to the radio regime, perhaps impacting the
observed \qtir$(z)$ behaviour. Investigating this possibility further
will be the subject of an upcoming paper.

\subsection{Systematics in the computation of radio luminosity\label{sub:discussion-Lrad}}

\begin{figure}
\includegraphics[bb=30bp 0bp 700bp 450bp,clip,scale=0.45]{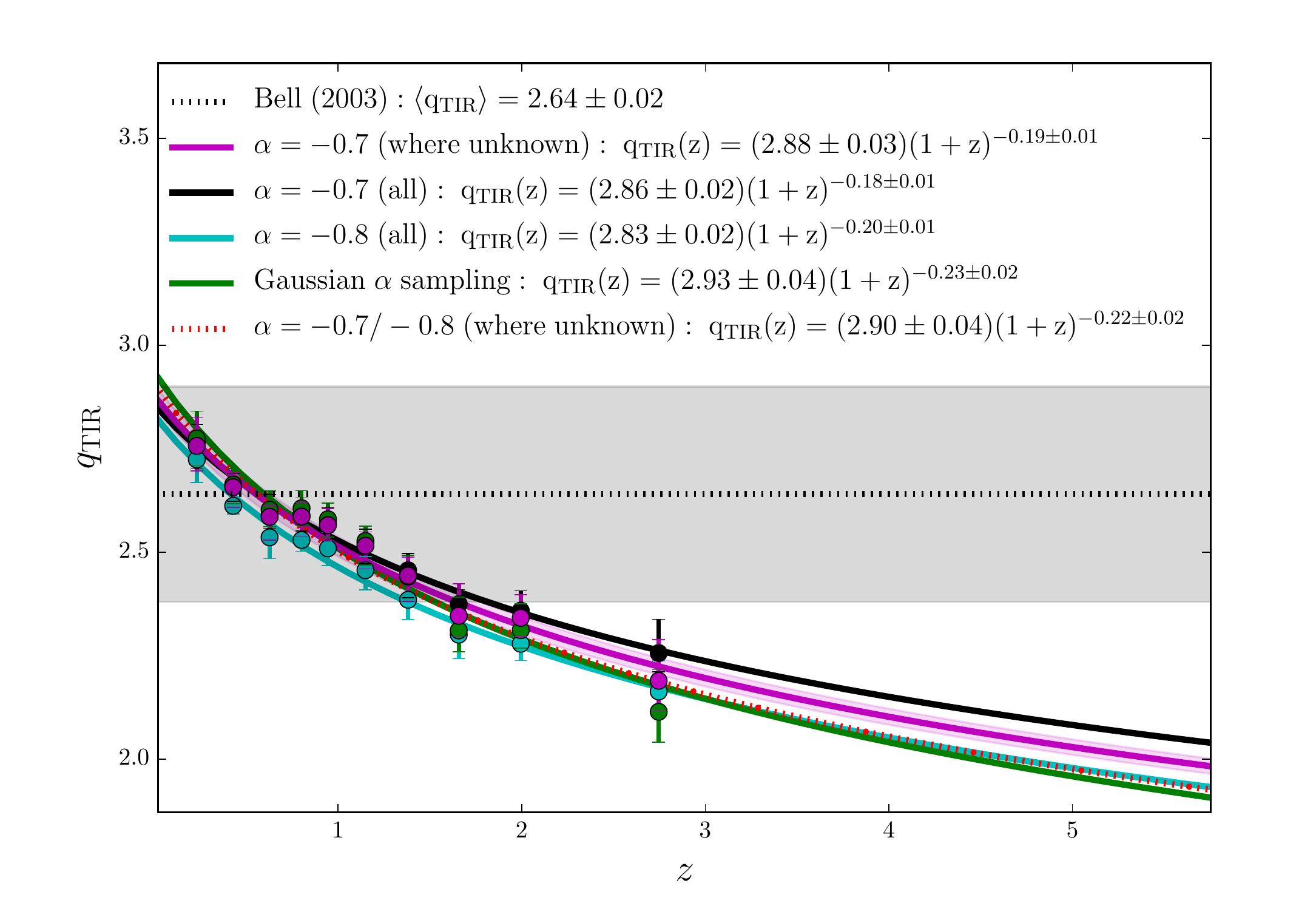}

\protect\protect\protect\protect\caption{Evolution of the IRRC found when using (i) the real spectral index,
where it is known, otherwise using $\alpha=-0.7$ (magenta; 1$\sigma$
uncertainty region shaded), (ii) a spectral index of $\alpha=-0.7$
for all sources (black), and (iii) a spectral index of $\alpha=-0.8$
for all sources (cyan). The green points and line show the result
of sampling $\alpha$ (where it is unknown) from a Gaussian distribution
with $\mu=-0.7$ and $\sigma=0.35$. The red dashed line shows the
use of $\alpha=-0.7$ (at $z<2$) and $-0.8$ (at $z>2$) where it
is unknown. \label{fig:q_vs_z_alphaanalytic}}
\end{figure}

\begin{figure}
\includegraphics[bb=60bp 0bp 350bp 400bp,clip,scale=0.8]{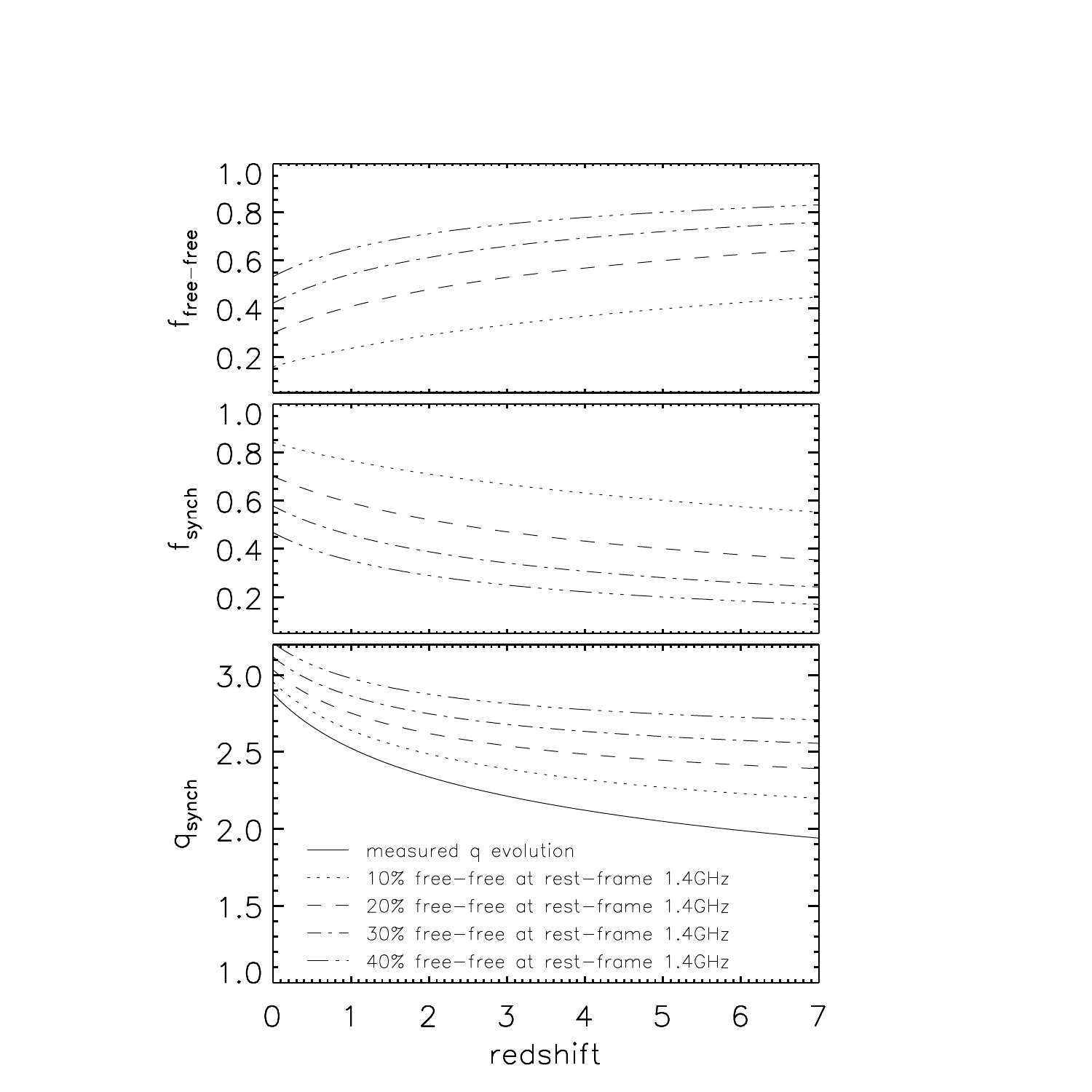}

\protect\protect\protect\protect\caption{Fractional contribution to 3\,GHz flux from free-free emission (top)
and synchrotron emission (middle) as a function of redshift, assuming
10, 20, 30, and 40 \% contributions of free-free emission at 1.4\,GHz
rest-frame frequency (see legend in bottom panel). The bottom panel
shows the power-law evolution of \qtir\ determined in Section \ref{sub:qtir}
(solid line), and the corrected evolution when the free-free emission
contribution is properly taken into account.\label{fig:freefree}}
\end{figure}

\begin{figure}
\includegraphics[scale=0.45]{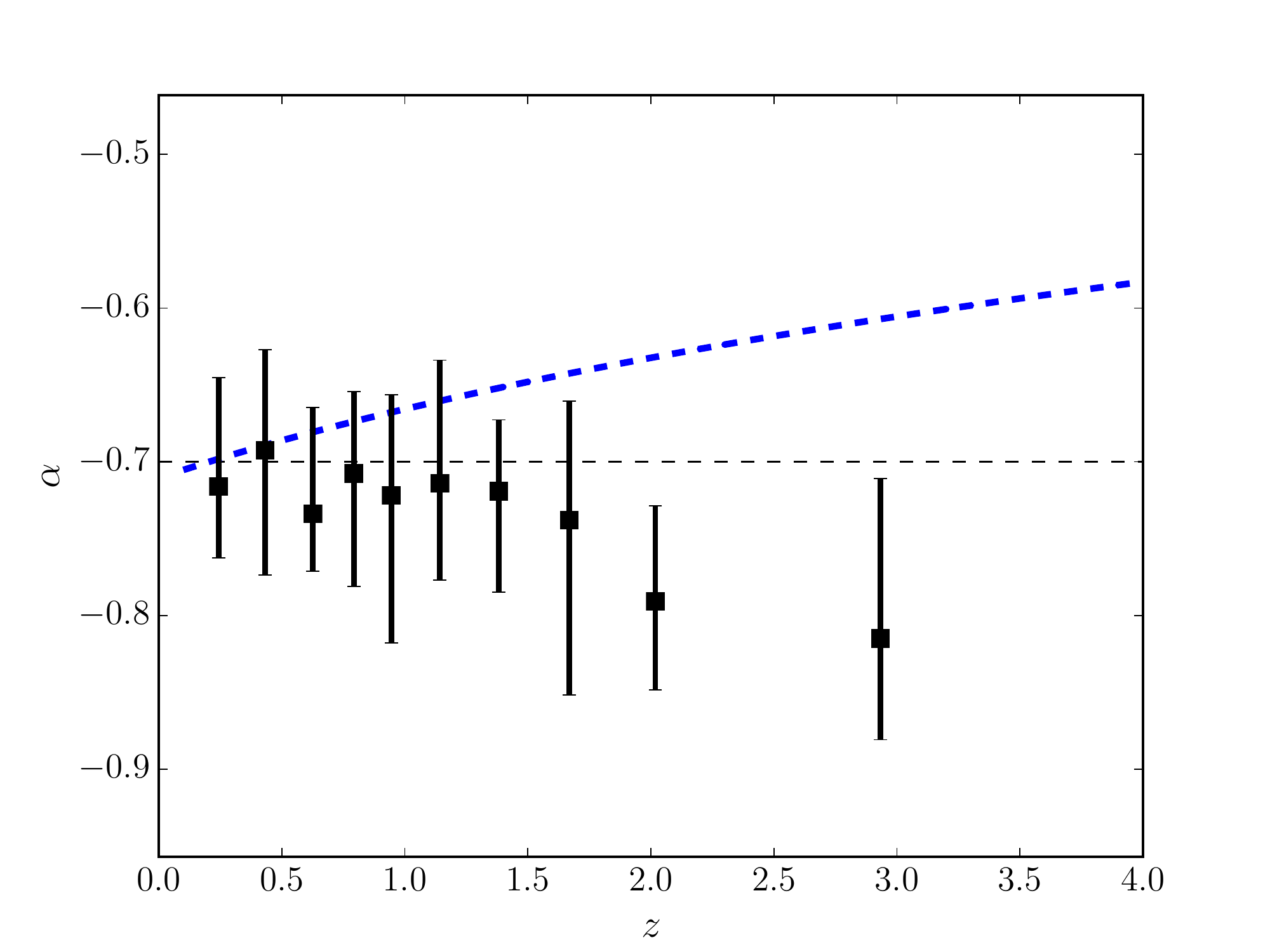}

\protect\protect\protect\protect\caption{Radio spectral index ($\alpha_{1.4\textrm{GHz}}^{3\textrm{GHz}}$)
of star-forming sources detected at 3\,GHz as a function of redshift.
As in Figure \ref{fig:alpha-vs-z}, the median values within each
redshift bin, derived from a single-censored survival analysis, are
shown by the black squares. The predicted evolution in the spectral
index due to the contamination of free-free emission based on the
M82 model of \citet{condon92} is shown by the thick, dashed blue
line. Our assumed value of $\alpha=-0.7$ for non-detections at 1.4\,GHz
is shown by the black dashed line.\label{fig:alpha}}
\end{figure}

In this section we investigate how the assumptions concerning the
exact spectral shape of the emission in the radio regime may affect
the derived IRRC.

\subsubsection{Influence of the radio spectral index}

We firstly examine the impact of the choice of the spectral index
($\alpha$) on the IRRC. As the IRRC is defined via a rest-frame 1.4\,GHz
luminosity (see ~\ref{eq:q}), which we here infer from the observed-frame
3\,GHz flux density (see ~\ref{eq:Lrad}), the choice of spectral
indices determines the $K$ corrections%
\footnote{We note that this is the case for any observing frequency even if,
for example, an observed 1.4~GHz flux density is used.%
}. As detailed in Section~\ref{sub:Radio-lums} we have made standard
assumptions, i.e.\ that the radio spectrum is a simple power law
($S_{\nu}\propto\nu^{\alpha}$). This is supported by the inferred
average spectral index of -0.7, approximately constant across redshift
(see Figure~\ref{fig:alpha-vs-z}), and consistent with that typically
found for star-forming galaxies, ($\alpha=-0.8$ to $-0.7$; e.g.~\citealt{condon92,kimball08,murphy09}).
We have therefore assumed $\alpha=-0.7$ for our 3\,GHz sources which
are undetected in the shallower 1.4\,GHz survey, while for the remainder
of the sources we have computed their spectral indices using the flux
densities at these two frequencies. From the expression for rest-frame
1.4\,GHz luminosity (Equation~\ref{eq:Lrad}) it follows that the
change in \qtir\ (Equation~\ref{eq:q}), when assuming two different
average spectral indices ($\alpha_{1}$ and $\alpha_{2}$, respectively),
is $\Delta$\qtir$(\alpha_{1},\alpha_{2})$$=-(\alpha_{1}-\alpha_{2})[\log(1+z)-\log(\frac{1.4}{3})]$.
For $\alpha_{1}=-0.7$, and $\alpha_{2}=-0.8$, $\Delta$\qtir$=-0.1\log(1+z)+0.033$.
This is illustrated in Figure~\ref{fig:q_vs_z_alphaanalytic} where
we show \qtir\ as a function of redshift derived i) using the measured
spectral index where it exists, otherwise setting $\alpha=-0.7$,
ii) with an assumed $\alpha=-0.7$ for all sources, and iii) with
an assumed $\alpha=-0.8$ for all sources. A change of 0.1 in the
assumed spectral index ($-0.7\rightarrow-0.8$) systematically lowers
\qtir\ and steepens the $(1+z)$ redshift dependence. Thus, the
choice of the average spectral index directly affects the normalisation,
as well as the derived trend with redshift of the IRRC. As discussed
in Section \ref{sub:Radio-lums}, the average spectral index of sources
in the two redshift bins at $z>2$ are consistent with $\alpha=-0.8$,
rather than $\alpha=-0.7$. We therefore also show in Figure \ref{fig:q_vs_z_alphaanalytic}
the \qtir\ trend found when assuming $\alpha=-0.7$ at $z<2$ and
$\alpha=-0.8$ at $z>2$ for sources where the spectral index is unknown.
Although slightly steeper, this is fully consistent with the use of
$\alpha=-0.7$ at all redshifts.

Finally, Figure \ref{fig:q_vs_z_alphaanalytic} also shows the results
of sampling the undefined spectral indices from a Gaussian distribution
centred at $\mu=-0.7$ and with a dispersion $\sigma=0.35$. This
is the distribution reported in \citet{smolcic17a} for all objects
detected in the VLA-COSMOS 3\,GHz Large Project. This $\alpha$ sampling
very slightly steepens the slope and increases the normalisation (albeit
within the uncertainties) due to the non-linear dependence of \lrad\
on $\alpha$ (see Equation \ref{eq:Lrad}).

\subsubsection{Influence of free-free contributions}

We next test whether the assumption of a simple power-law is a realistic
description of the spectral energy distribution in the radio regime.
Synchrotron emission is a major component of typical radio SEDs for
star-forming galaxies at rest-frequencies of $\sim1-20$\,GHz. At
higher frequencies, free-free (Bremsstrahlung) emission begins to
contribute substantially (see e.g.~Figure~1 in \citealt{condon92}).
Both emission processes can be described as power-law radio spectra
($S_{\nu}\propto\nu^{\alpha}$), with a spectral index of $-0.8$
(synchrotron emission), and $-0.1$ (free-free emission). For low
redshift galaxies, the observing frequencies probe the rest-frame
part of the spectrum dominated by the synchrotron emission. However,
towards higher redshifts the free-free contributions at rest-frame
frequencies become increasingly significant.

In Figure \ref{fig:freefree} we show the expected fractional contribution
of free-free emission as a function of redshift, assuming various
(10-40\%) fractional contributions of free-free emission at 1.4\,GHz
rest-frame. The corresponding synchrotron fractions are also shown
as a function of redshift.

The bottom panel of Figure \ref{fig:freefree} shows \qtir$(z)$
if we exclude the free-free contribution and calculate \qtir\ using
only the synchrotron contribution to the total observed radio emission.
The slope of \qtir$(z)$ is flatter, however a declining trend with
redshift is still observed when a 10\% contribution of free-free emission
at rest-frame 1.4~GHz frequency is assumed (consistent with \citealt{condon92,murphy09}).
However, the local \qtir\ value is then at the high end of that
locally derived by numerous studies (e.g.~\citealt{bell03}).

Examining the variation of the spectral index as a function of redshift
may also provide information on the extent of the free-free contribution.
If we again assume a simplistic radio SED with $\alpha=-0.8$ for
synchrotron emission and $\alpha=-0.1$ for free-free emission, then
we expect a flattening of the average observed radio spectral index
towards higher redshifts. A higher rest-frame frequency is sampled
at higher redshifts, given a fixed observing frequency. Since the
fractional contribution of free-free emission is larger at higher
frequencies, the measured total flux will be larger and hence the
spectral index flatter.

Assuming a 10\% contribution of free-free emission to the total radio
flux density at rest-frame 1.4\,GHz, we find that the change of the
average spectral index amounts to $\Delta\alpha(z)=\alpha(z=4.0)-\alpha(z=0.2)=0.11$
only. We note that the average spectral index is, under these assumptions,
consistent with the local average, $\alpha(z=0.2)=-0.7$ value inferred
using the real data. If we assume free-free emission contributions
to the total radio spectrum at rest-frame 1.4\,GHz frequency of 20\%,
30\%, and 40\%, we infer an increase (i.e.~flattening) of the observed
spectral index of only $\Delta\alpha=0.17$ (albeit with a steeper
local spectral index than inferred for the real data). However, the
flattening of the average radio spectral index expected under the
given assumptions is not supported by our data, as can be seen in
Figure~\ref{fig:alpha}.

The general conclusion is that the fractional contribution of free-free
emission to the observed radio spectrum with standard, simple assumptions
is inconsistent with the derived decreasing trend of \qtir\ with
increasing redshift. This suggests a more complex radio SED for star-forming
galaxies, compared to the usual assumptions of a superposition of
$\alpha=-0.8$ and $\alpha=-0.1$ power-law synchrotron and free-free
spectra, respectively, such that at rest-frame 1.4\,GHz the free-free
contribution amounts to 10\% of the total radio emission (e.g.~\citealt{condon92,yun02,bell03,murphy09,galvin16}).

\subsubsection{Comparison with local (U)LIRGs}

\begin{figure}
\includegraphics[scale=0.45]{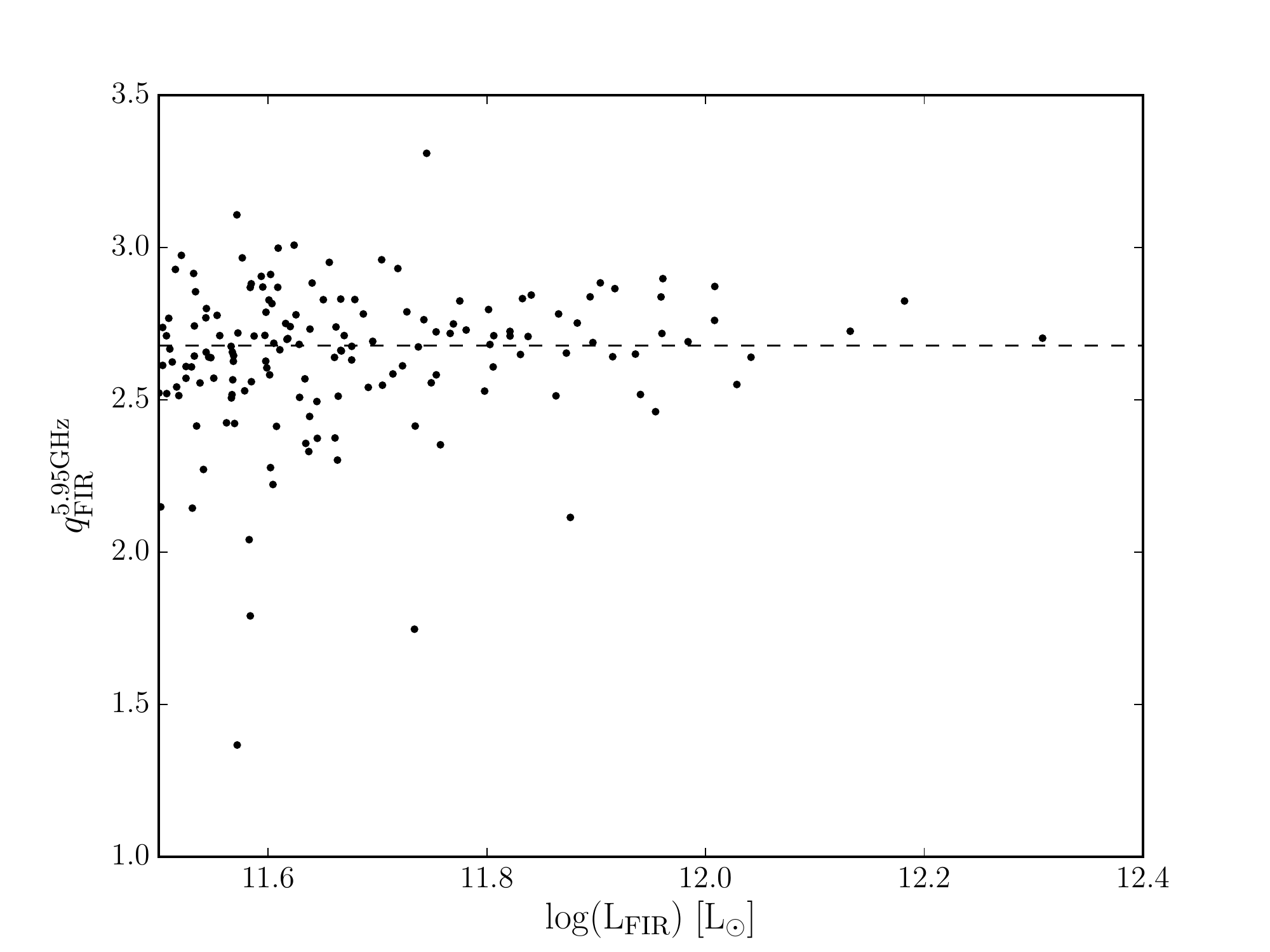}

\protect\protect\protect\protect\caption{IRRC defined at 5.95\,GHz versus $L_{{\rm FIR}}$ for star-forming
objects in our sample at $0.9<z<1.1$. The dashed line indicates the
median value: $q_{{\rm FIR}}^{5.95{\rm GHz}}=2.68\pm0.02$.\label{fig:q6GHz}}
\end{figure}

The radio SED for star-forming galaxies was studied by \citet{leroy11}
who obtained VLA observations of local ($z\sim0$) (ultra-) luminous
infrared galaxies - (U)LIRGs - in C-band (5.95\,GHz). They calculated
the IRRC in this band and found a median value of $q_{{\rm FIR}}^{5.95{\rm GHz}}=2.8$
with a scatter of 0.16 dex. At $z\sim1$, rest-frame 5.95\,GHz corresponds
to observed-frame 3\,GHz. This means that we can use our 3\,GHz
data to calculate $q_{{\rm FIR}}^{5.95{\rm GHz}}$ with no, or very
little, $K$ correction required for objects in our sample at $z\sim1$.
Figure \ref{fig:q6GHz} shows $q_{{\rm FIR}}^{5.95{\rm GHz}}$ versus
$L_{{\rm FIR}}$ for objects in our sample at $0.9<z<1.1$ and with
$\log(L_{{\rm FIR}})>11.5L_{\odot}$ for the sake of completeness
and a fair comparison (although we note that this restricts us to
a luminosity range of $\sim1$dex). We find a median $q_{{\rm FIR}}^{5.95{\rm GHz}}=2.68\pm0.02$
with a scatter of 0.24 dex. The $L_{{\rm FIR}}$ range of these objects
matches closely with the (U)LIRGs sample of \citet{leroy11}. Therefore,
we can directly compare the two samples. We find that minimising $K$
corrections in the radio band flattens the observed trend of decreasing
\qtir\ with increasing redshift. The inferred $q_{{\rm FIR}}^{5.95{\rm GHz}}$
value at $z=1$ is consistent with a trend $\propto(1+z)^{-0.06}$
(rather than with $q_{{\rm FIR}}(z)\propto(1+z)^{-0.21}$ as derived
in Section \ref{sub:lit-comparison}). This suggests that the observed
redshift trend of \qtir\ may be at least partially attributable
to uncertainties in the $K$ corrections applied to the radio flux.
Therefore, further investigations into the radio spectra of various
star-forming galaxy populations are required for robust determinations
of $K$ corrections in the radio regime, having particular relevance
for high-redshift star-forming galaxies.

\subsection{Other physical factors}

Along with uncertainty in the radio SED shape and the possible contribution
from AGN, it is possible that other physical mechanisms could be driving
a decrease in \qtir\ towards higher redshifts. While a thorough
investigation of these is beyond the scope of this paper, we nonetheless
mention several mechanisms here. One possible driver of an evolving
\qtir$(z)$ is the changing magnetic field properties of galaxies.
An increasing magnetic field strength would increase the flux of synchrotron
radiation and thereby decrease the measured \qtir. While galaxy-scale
magnetic fields are thought to build up over time (e.g.~\citealp{beck96})
perhaps from turbulent seed fields (\citealp{Arshakian09}), the mean
magnetic field strength in a galaxy undergoing a global starburst
may be elevated. \citet{tabatabaei17} argue that the amplification
of magnetic fields within star-forming regions in galaxies with high
SFRs could result in a decrease of the infrared-radio correlation.
Such a decrease may be stronger at higher redshifts due to the detection
bias towards objects with higher SFRs.

It is also thought that major mergers of galaxies can enhance synchrotron
emission through various processes and thus result in a decreased
measurement of \qtir. For example, \citet{kotarba10} performed a
magnetohydrodynamical simulation of NGC 4038 and NGC 4039 (the Antennae
galaxies) and found evidence for amplification of magnetic fields
within merging systems due to compression and shear flows. As discussed
above, an increased magnetic field strength would increase the synchrotron
emission from pre-existing cosmic rays. \citet{murphy13} studied
a sample of nearby steep-spectrum infrared-bright starburst galaxies
and argue that gas bridges between the interacting taffy-like systems
could be the site of enhanced synchrotron radiation which is not related
to star formation. Furthermore, \citet{donevski15} argue that, in
addition to the effects of enhanced magnetic fields, shocks generated
by galactic interactions will accelerate electrons and thus further
boost synchrotron emission.

The timescale for merger-enhanced infrared emission (due to shock-heating
of gas and dust) is expected to be on the order of $\sim$10 Myr,
followed by the enhanced synchrotron emission phase which is expected
to last from hundreds of Myrs up to a Gyr \citep{donevski15}. If
this is the case, then it is statistically more likely that a flux-limited
sample contains more galaxies in the phase of synchrotron boosting
(Prodanovi\'{c}; private communication). Thus, it is possible that
an increasing major merger fraction with redshift, such as that presented
by \citet{conselice14} to $z\sim3$, could partially explain a decreasing
\qtir$(z)$.

We note that this is not a comprehensive list of the many physical
processes which could be driving an evolving \qtir$(z)$ and that
a number of competing mechanisms, such as inverse Compton energy losses
towards higher redshifts (e.g. \citealp{murphy09}), could also be
at play.

\subsection{Radio as a star-formation rate tracer}

We have determined that \qtir\ decreases with increasing redshift,
consistent with previous results in the literature (e.g.~\citealt{ivison10,sargent10a,magnelli15}).
In Section \ref{sub:discussion-Lrad}, we have shown that this trend
may partly be due to uncertainties in the $K$ correction in the radio
due to the overly simplistic assumptions that the radio spectrum can
be well-described by a simple power-law. Nevertheless, regardless
of the origin of the observed trend, we can make use of it to recalibrate
radio luminosity as a SFR tracer as a function of redshift.

In the local Universe, 1.4\,GHz rest-frame radio luminosity is anchored
to the SFR via the \qtir\ parameter (e.g.~\citealt{condon92,yun01,bell03}).
Following \citet{yun01}, we make use of the Kennicutt (1998) calibration
for total IR luminosity based SFR:

\begin{equation}
\mathrm{SFR~[M_{\odot}/yr]}=f_{\mathrm{IMF}}10^{-10}L_{\mathrm{IR}}~\mathrm{[L_{\odot}]},
\end{equation}

where SFR is the star formation rate in units of M$_{\odot}$/yr,
$f_{\mathrm{IMF}}$ is a factor accounting for the assumed initial
mass function (IMF, $f_{\mathrm{IMF}}=1$ for a Chabrier IMF, $f_{\mathrm{IMF}}=1.7$
for a Salpeter IMF), and $L_{\mathrm{IR}}$ is the total IR luminosity
in units of Solar luminosities. Relating the SFR to the rest-frame
1.4\,GHz luminosity through Equation \ref{eq:q} and accounting for
the redshift and radio spectral index dependences then yields:

\begin{equation}
\mathrm{SFR~[M_{\odot}/yr]}=f_{\mathrm{IMF}}10^{-24}10^{q_{\mathrm{TIR}}(z,\alpha)}L_{\mathrm{1.4GHz}}~\mathrm{[W/Hz]},
\end{equation}

where

\[
q_{\mathrm{TIR}}(z)=\begin{cases}
(2.88\mbox{\ensuremath{\pm}}0.03)(1+z)^{-0.19\pm0.01} & \text{for}\quad\left\langle \alpha\right\rangle =-0.7\\
(2.85\mbox{\ensuremath{\pm}}0.03)(1+z)^{-0.22\pm0.01} & \text{for}\quad\left\langle \alpha\right\rangle =-0.8
\end{cases},
\]
where $\left\langle \alpha\right\rangle $ is the average assumed
spectral index of the star-forming galaxy population, and

\begin{equation}
L_{1.4{\rm GHz}}=L_{1.4{\rm GHz}}(z,\alpha)=\frac{4\pi D_{L}^{2}}{(1+z)^{\alpha+1}}(\frac{1.4}{\nu_{\mathrm{obs}}})^{\alpha}S_{\nu_{\mathrm{obs}}},\label{eq:Lradgeneral}
\end{equation}

where $\nu_{\mathrm{obs}}$ is the observing frequency in units of
GHz, here tested and verified for $\nu_{\mathrm{obs}}=1.4$ and $3$\,GHz,
and $\alpha=-0.7$, and $-0.8$. It is important to note that the
above is valid only for samples of star-forming galaxies selected
similarly to those studied here and under the assumptions: (i) of
a luminosity-independent IRRC, (ii) of simple $K$ corrections of
the radio spectrum ($S_{\nu}\propto\nu^{\alpha}$) as presented in
Equation \ref{eq:Lradgeneral}, and (iii) that the infrared luminosity
accurately traces the SFR with redshift.

\section{Conclusions}

We use the new, sensitive VLA-COSMOS 3\,GHz Large Project and infrared
data from Herschel and Spitzer to push studies of the infrared radio
correlation (IRRC) out to $z\sim6$ over the 2\,deg$^{2}$ COSMOS
field. The excellent sensitivity of the 3\,GHz data allows us to
directly detect objects down to the micro-Jansky regime. We detect
7,729 sources in the 3\,GHz data with optical counterparts and redshifts
available in the COSMOS database. We identify 8,458 sources detected
in the Herschel PEP and HerMES surveys with counterparts in Spitzer
MIPS 24\,$\mu$m data and in the optical. Our final sample, jointly-selected
in both the radio and infrared, consists of 12,333 unique objects.

We take advantage of the plethora of high-quality multiwavelength
data available in the COSMOS field, as well as our ability to perform
a multi-component SED fitting process, to separate our sample into
(non-active) star-forming galaxies, moderate-to-high radiative luminosity
AGN (HLAGN) and low-to-moderate radiative luminosity AGN (MLAGN).
We study the IRRC for each of these populations separately.

We examine the behaviour of the IRRC, characterised by the \qtir\
parameter, as a function of redshift using a doubly-censored survival
analysis to account for non-detections in the radio or infrared along
with a bootstrap approach to incorporate measurement errors. A slight,
but statistically significant, trend of \qtir\ with redshift is
found for the population of star-forming galaxies: $q_{{\rm TIR}}(z)=(2.88\pm0.03)(1+z)^{-0.19\pm0.01}$.
This is in good agreement with several other results from the literature,
although is biased slightly high compared to studies of the local
Universe. To examine biases introduced by the sensitivity limits of
our data, we perform various tests incorporating these local measurements,
and/or ignoring non-detections. In all cases we find a statistically-significant
decrease of\textbf{ }\qtir\ with increasing redshift, with the slope
(i.e.~$(1+z)$ exponent) ranging between -0.20 and -0.09.

When examined separately, we find that AGN have \qtir\ measurements
biased towards lower values, suggesting that radio wavelengths are
more likely than the infrared to be influenced by emission from active
processes. It is possible that AGN contributions only to the radio
regime could be influencing (i.e.~steepening) the observed \qtir$(z)$
trend, particularly if this occurs in an appreciable fraction of star-forming
host galaxies.

We find that the choice of radio spectral index used for the $K$
correction of the 3\,GHz flux can influence both the shape and normalisation
of the \qtir$(z)$. The increasing contribution of free-free emission
towards higher radio frequencies may also influence the redshift trend,
however our results are inconsistent with a typical (M82-based) model
of the radio SED. We conclude that a better understanding of the SED
of star-forming galaxies is needed for a comprehensive physical interpretation
of the apparent redshift evolution of the IRRC. Other physical mechanisms
which could potentially drive a decreasing \qtir$(z)$ include changing
galaxy magnetic field strengths and major merger fractions.

Finally, we present a redshift-dependent relation between rest-frame
1.4\,GHz luminosity and star formation rate. 
\begin{acknowledgements}
We thank the anonymous referee for useful comments which have helped
to improve this paper. We also thank Tijana Prodanovi\'{c} for valuable
discussions. This research was funded by the European Unions Seventh
Frame-work programme under grant agreement 337595 (ERC Starting Grant,
`CoSMass'). This research was supported by the Munich Institute for
Astro- and Particle Physics (MIAPP) of the DFG cluster of excellence
\textquotedbl{}Origin and Structure of the Universe\textquotedbl{}.
NB acknowledges the European Unions Seventh Framework programme under
grant agreement 333654 (CIG, AGN feedback). MB and P.\,Ciliegi acknowledge
support from the PRIN-INAF 2014. AK acknowledges support by the Collaborative
Research Council 956, sub-project A1, funded by the Deutsche Forschungsgemeinschaft
(DFG). MTS acknowledges support from a Royal Society Leverhulme Trust
Senior Research Fellowship. Support for B.M. was provided by the DFG
priority programme 1573 \textquotedbl{}The physics of the interstellar
medium\textquotedbl{}. M.A. acknowledges partial support from FONDECYT
through grant 1140099. 
\end{acknowledgements}

 \bibliographystyle{aa}
\bibliography{IR-radio_bib}

\end{document}